\documentclass[lettersize,journal]{IEEEtran}
\usepackage{amsmath,amsfonts}
\usepackage{algorithmic}
\usepackage{array}
\usepackage[caption=false,font=normalsize,labelfont=sf,textfont=sf]{subfig}
\usepackage{textcomp}
\usepackage{stfloats}
\usepackage{url}
\usepackage{verbatim}
\usepackage{graphicx}
\usepackage{amsmath,amssymb,amsfonts}
    \usepackage{graphicx}
    \usepackage[T1]{fontenc} 
    \usepackage[cmintegrals]{newtxmath}
    \usepackage{textcomp}
    \usepackage{bm}
    \usepackage{nicematrix}
    \usepackage{tikz}
    \usepackage{tabu}
    \usepackage[lined,boxed,ruled]{algorithm2e}
    \usepackage{multirow}
    \usepackage{booktabs}
    \usepackage{stfloats}
    \usepackage{times}
    \usepackage{epsfig}
    \usepackage{bm}
    \usepackage{color}
    \usepackage{arydshln}
    \usepackage{mathrsfs}
    \usepackage{subcaption}
    \usepackage{tabularx}
\usepackage[letterpaper=true,colorlinks,bookmarks=false]{hyperref}

\ExplSyntaxOn

\keys_define:nn { MyRule }
  {
    color .tl_set:N = \l__MyRule_color_tl , 
    color .value_required:n = true ,
    width .dim_set:N = \l__MyRule_width_dim ,
    width .value_required:n = true ,
    width .initial:n = \arrayrulewidth ,
    style .value_required:n = true ,
    style .tl_set:N = \l__MyRule_style_tl
  }

\NewDocumentCommand { \MyRule } { O { } D ( ) { } m }
  {
    \exp_args:Nx \__MyRule_i:nnnn 
    { \int_use:c { c@iRow } } { #1 } { #2 } { #3 } 
  }

\cs_new_protected:Nn \__MyRule_i:nnnn
  {
    \tl_gput_right:Nn \g_nicematrix_code_after_tl 
      { \__MyRule_ii:nnnn { #1 } { #2 } { #3 } { #4 } }
    \peek_remove_spaces:n { }
  }

\cs_new_protected:Nn \__MyRule_ii:nnnn
  {
    \group_begin:
    \keys_set:nn { MyRule } { #2 }
    \__MyRule_iii:w #4 \q_stop
    \begin { tikzpicture }
    \pgfsetlinewidth { \l__MyRule_width_dim }
    \tl_if_empty:NF \l__MyRule_color_tl    
      { \pgfsetstrokecolor { \l__MyRule_color_tl } }
    \tl_if_in:nnTF { #3 } { l }
      { \dim_set:Nn \l_tmpa_dim { 1 mm } }
      { \dim_zero:N \l_tmpa_dim }
    \tl_if_in:nnTF { #3 } { r }
      { \dim_set:Nn \l_tmpa_dim { 1 mm } }
      { \dim_zero:N \l_tmpa_dim }
    \tl_if_empty:NF \l__MyRule_line_style_tl
      { \tikzset { every~path/.style = \l__MyRule_style_tl  } }
    \draw ([xshift=\l_tmpa_dim,yshift=-\l__MyRule_width_dim/2] #1 
             -| \int_use:N \l_tmpa_int ) --
          ([xshift=-\l_tmpa_dim,yshift=-\l__MyRule_width_dim/2] #1 
             -| \int_use:N \l_tmpb_int )  ;
    \end { tikzpicture }
    \group_end:
  }

\cs_new_protected:Npn \__MyRule_iii:w #1 - #2 \q_stop
  {
    \int_set:Nn \l_tmpa_int { #1 }
    \int_set:Nn \l_tmpb_int { #2 + 1 }
  }

\ExplSyntaxOff

\def\BibTeX{{\rm B\kern-.05em{\sc i\kern-.025em b}\kern-.08em
    T\kern-.1667em\lower.7ex\hbox{E}\kern-.125emX}}
\usepackage{balance}
\begin{document}
\title{Structure and Intensity Unbiased Translation for 2D Medical Image Segmentation}
\author{Tianyang Zhang, Shaoming Zheng,
        Jun Cheng, Xi Jia,  Joseph Bartlett, Xinxing Cheng,
         Zhaowen Qiu,\\
        Huazhu Fu,
        Jiang Liu,  Aleš Leonardis, and Jinming Duan
\thanks{\IEEEcompsocthanksitem T. Zhang, X. Jia, J. Bartlett, X. Cheng A. Leonardis and J. Duan are with the University of Birmingham, Birmingham, UK. 
\IEEEcompsocthanksitem J. Bartlett is also with the University of Melbourne, Victoria, Australia
\IEEEcompsocthanksitem A. Leonardis and J. Duan (email: j.duan@bham.ac.uk) are also Fellows of the Alan Turing Institute, London. 
\IEEEcompsocthanksitem S. Zheng is with University of Pennsylvania, USA. 
\IEEEcompsocthanksitem J. Cheng is with Institute for Infocomm Research, A*STAR, Singapore. (e-mail: sam.j.cheng@gmail.com). 
\IEEEcompsocthanksitem H. Fu is with Institute of High Performance Computing, A*STAR, Singapore.
\IEEEcompsocthanksitem Z. Qiu is with the NorthEast Forestry University, Harbin, China. J. Liu is with the Southern University of Science and Technology, Shenzhen, China. 
\IEEEcompsocthanksitem The corresponding authors are Jun Cheng  and Jinming Duan.}}

\markboth{Journal of \LaTeX\ Class Files,~Vol.~18, No.~9, September~2020}%
{How to Use the IEEEtran \LaTeX \ Templates}

\maketitle

\begin{abstract}
Data distribution gaps often pose significant challenges to the use of deep segmentation models. However, retraining models for each distribution is expensive and time-consuming. In clinical contexts, device-embedded algorithms and networks, typically unretrainable and unaccessable post-manufacture, exacerbate this issue. Generative translation methods offer a solution to mitigate the gap by transferring data across domains.  However, existing methods mainly focus on intensity distributions while ignoring the gaps due to structure disparities.  In this paper, we formulate a new image-to-image translation task to reduce structural gaps. We propose a simple, yet powerful Structure-Unbiased Adversarial (SUA) network which accounts for both intensity and structural differences between the training and test sets for segmentation. It consists of a spatial transformation block followed by an intensity distribution rendering module. The spatial transformation block is proposed to reduce the structural gaps between the two images. The intensity distribution rendering module then renders the deformed structure to an image with the target intensity distribution. Experimental results show that the proposed SUA method has the capability to transfer both intensity distribution and structural content between multiple pairs of datasets and is superior to prior arts in closing the gaps for improving segmentation.
\end{abstract}


\section{Introduction}
\label{sec:introduction}
\IEEEPARstart{M}{edical} imagesegmentation~\cite{gu2019net, chen2019synergistic} has been a hot topic in the last few decades, in particular, deep learning based techniques \cite{litjens2017survey,shen2017deep} have drawn lots of attention. These methods often assume that the training and test data follow the same distribution.
 However,   domain gaps often exist between data from different sources, \emph{e.g.}, the data from different hospitals or clinic centers is often captured by different machines with different  settings. In addition, as imaging technology continues to progress, old machines become outdated and are therefore replaced by their modern counterparts. Therefore, even the new data and the accumulated data from the same place have different distributions. When it comes to performing inference using deep learning techniques, such differences between the  training and inference distributions can degrade performance significantly. However, it is clear that recollecting and labelling the required training data for each distribution is hugely expensive and time-consuming. Therefore, the effective use of labelled data from previous devices or settings is vital.
\begin{figure}[!t]
  \includegraphics[width=\linewidth]{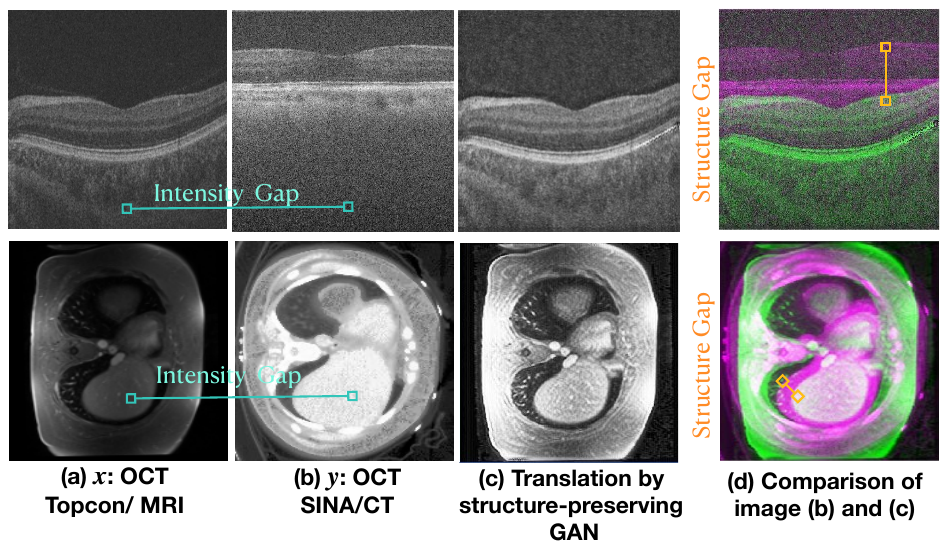}
  \caption{Images illustrate the domain shift problem, with green and orange lines showing differences in intensity distribution and structure compared to the target image. The comparison images on the right side highlight that structural gaps, e.g. curve, position and anatomy, are challenging to mitigate through translation methods that keep the structure unchanged. Specifically, $\bm x$ represents a sample from the source dataset, while $\bm y$ represents a sample from the target dataset. The purple and green regions indicate areas where the representation in (b) differs from (c), and where the representation in (c) differs from (b), respectively.}
  \label{OCT_demos}
\end{figure}	

	To solve this issue, transfer learning methods~\cite{DECAF,duan2012domain,JAN},  such as unsupervised domain adaptation (UDA)~\cite{chen2019synergistic,dou2018unsupervised,jiang2020psigan,carlucci2020multidial},  are possible ways to map the data into a different space such that the domain gaps between the source domain and target domain are minimized. A limitation of such approaches is that they often require the model to be retrained or its latent features to be accessed. However, networks and labels compiled to software can not be touched in clinical applications.  Recently, generative adversarial models have been proposed to tackle this problem by transferring the intensity distributions from source to target domain and reducing the domain gaps. In this task, the input image is transferred (adapted) and then tested with a model trained on the original labelled data. Specifically, this task has been defined in~\cite{gadermayr2019generative} and is considered to be different from  UDA or other similar problems.  

\begin{figure*}[!t]
\includegraphics[width=\textwidth]{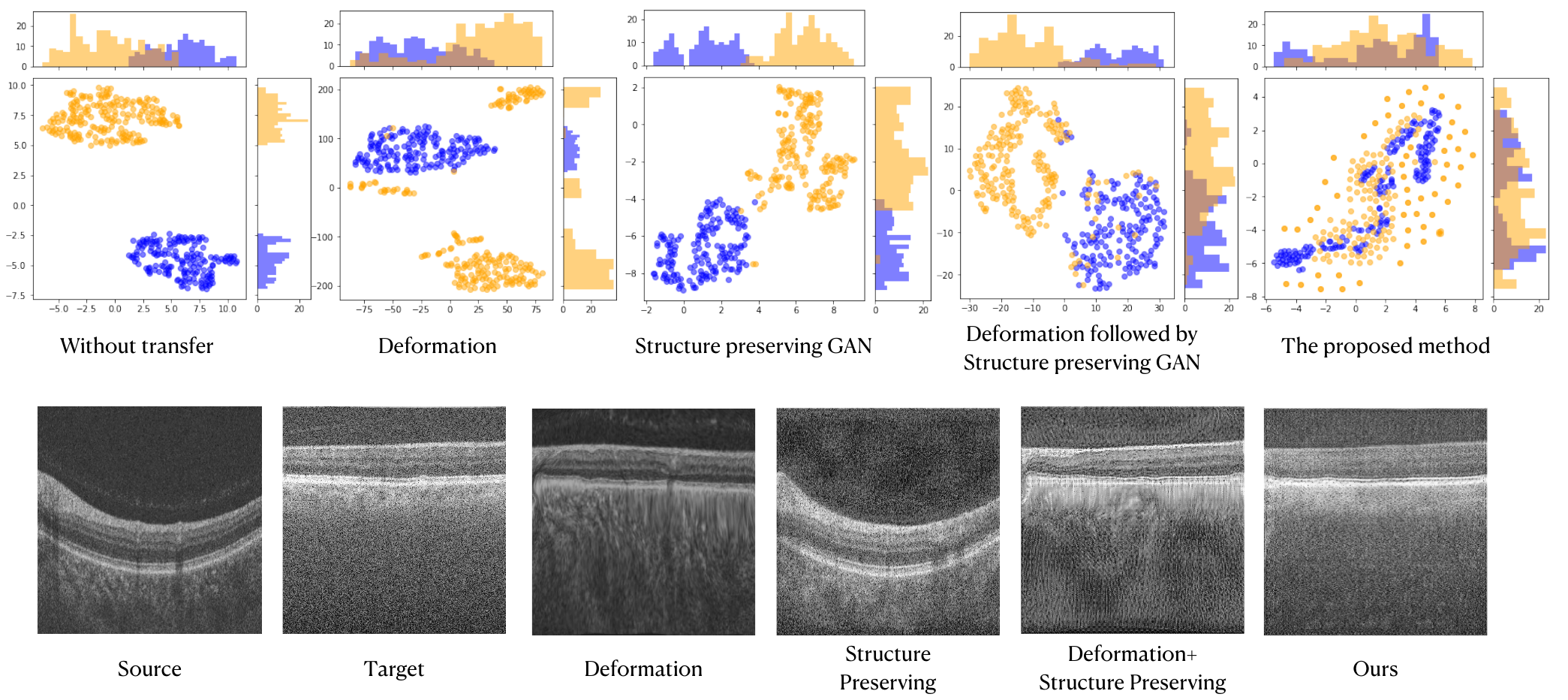}
  \caption{\textcolor{black}{Illustrations of domain shift issues using T-SNE. Yellow and blue dots indicate images from two different domains. From left to right, figures are plotted based on: (a) original data points; (b) data points transformed by spatial deformation; (c) data points transferred by structure-preserving GAN; (d) data points transformed first by spatial deformation followed by structure-preserving GAN; and (e) data points transferred by our proposed SUA method. The second row shows some examples.}}
  \label{tsne}
\end{figure*}

 Previously, Chen et.~al~\cite{chen2019unsupervised} proposed a MUNIT based model to transfer the intensity distribution of the source dataset to that of the target dataset by maintaining the content in the latent space.  Zhang et.~al~\cite{zhang2019noise} proposed to generate new images by maintaining the main edges in the images. However, these   methods overlooked the differences in structure statistics. In fact, the structures captured using various imaging techniques can be different due to the nature of   imaging principles.  For instance,   the curvatures of the   objects of interest in images are largely affected by scales.   Fig.~\ref{OCT_demos}  shows two OCT images (a) and (b) from different manufacturers. As shown in sub-image (a), not only structures but also intensities are different from the target (b).  Therefore, we have a task with contradictive objectives: on the one hand, we hope to maintain the contents/edges in the generated data for the subsequent analysis. It leads to the propagation of the structural gap. On the other hand, we hope to eliminate the structural gap as the generated and   real data are supposed to be indistinguishable to the discriminator. Overall, we present the translation problem to closing the existing domain gaps of structure and intensity distribution and improve the segmentation performance. For this segmentation problem under consideration, a model is meticulously trained using target domain images alongside their associated masks. The trained model is subsequently tested on the samples translated from source domain to target domain.
 
\begin{figure}[h]
  \includegraphics[width=\linewidth]{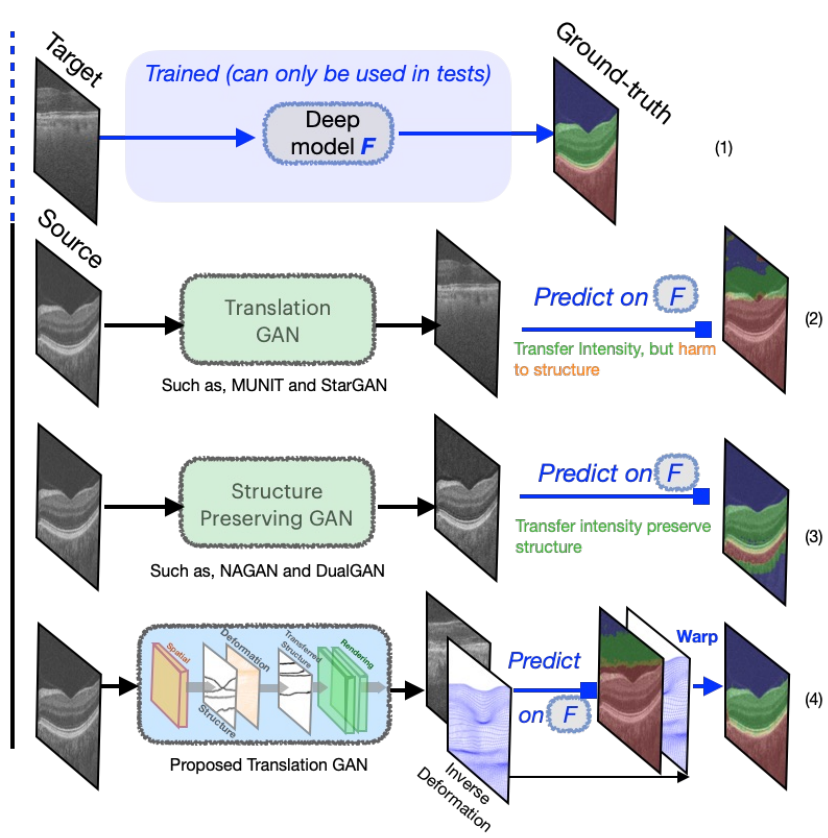}
  \caption{Illustration of pipelines of translation methods for segmentation: (1) Trained segmentation model without domain adaptation; (2) Translation GAN; (3) Structure preserving GANs, and (4) our GAN with learnable deformation (more details in Fig.~\ref{pipline}).}
  \label{examples}
\end{figure}

 In this paper, we propose a simple yet important structure unbiased adversarial approach (SUA) to overcome the above issue. It extracts the main structures from the image and translates them separately  such that both structures and intensities gaps are reduced. With this translation, it is expected that the performance of models trained on the training data (target) will improve when tested on data (source) from a different distribution. Finally, we also compute an inverse deformation field to warp the segmentation result back to its original domain, which is an essential step to obtain the segmentation for the original images.  Note that our method is different from a spatial deformation followed by a structure-preserving GAN. \textcolor{black}{To better understand  the motivation of the proposed method,  we use T-SNE \cite{tsne08}  to visualize OCT images after different processing in Fig.~\ref{tsne}. }  As shown, there is a large gap between the  data from the two domains indicated by the yellow and blue dots. Although a deformation could reduce the gap, the gap due to intensity distributions remain, as shown in (b). Similarly, (c) illustrates that after applying a structure preserving GAN to the images they can still be distinguished. Whilst applying a deformation followed by a structure preserving GAN is able to reduce the gap, the deformation causes significant texture distortions. Such distortions cannot be removed by the structure-preserving translation GAN, as is described in Section~\ref{combination}. Additionally, a brief illustration of the proposed method compared with other translation methods is shown in Fig.~\ref{examples}.

The main contributions of the paper are summarized as follows:
	\begin{itemize}
		\item We find that the domain gaps exist in not only the intensity distributions but also in the image structures. Then we identify  an important image-to-image translation problem  which helps to reduce domain gaps associated with both image structure and intensity distributions. By taking this advantage,  we demonstrate the effectiveness of our method for medical image segmentation.\\
  
		\item We propose a novel SUA network to tackle domain gaps in segmentation problem by   a novel image-to-image translation strategy. Specifically, the proposed method has the capability to translate images by reducing the structural and intensity gaps with a spatial transformation block and a structure rendering mapping respectively.\\ 
  
		\item Extensive experiments on two retinal  OCT datasets, a chest CT \& MRI paired dataset and two cardiac datasets show that the proposed method is able to transfer both  structure shapes and intensity distributions effectively with improved subsequent segmentation results.
	\end{itemize}


\section{Related works}
\label{sec:related}

\subsection{Generative Adversarial Networks (GANs)} GANs\cite{goodfellow2014generative,denton2015deep}
are originally proposed to generate images from  random inputs  in an unconditional manner. They contain a generator and a discriminator; the generator aims to output samples to be indistinguishable from the training samples while the discriminator tries to differentiate them. 
	Recently, many conditions have been proposed and integrated into GANs for various applications   such as  image segmentation \cite{isola2017image, luc2016semantic,dou2018unsupervised},  image synthesis \cite{kazemi2018facial},  super-resolution \cite{ledig2017photo},   image reconstruction\cite{wolterink2017deep,wolterink2017generative,wang20183d}, medical image translation \cite{zhao2018synthesizing, iqbal2018generative, armanious2020medgan, armanious2019unsupervised} and text-to-images synthesis \cite{reed2016generative}. Such conditional GANs are generally divided into  paired image-to-image translation and unpaired image-to-image translation.

\subsection{Paired Image-to-image Translation}
 Paired image-to-image translation methods are trained on a paired dataset to obtain a mapping which converts an image from one domain to another \cite{isola2017image, tripathy2018learning}. Earlier, many paired image-to-image translation algorithms have been proposed  for various   tasks, e.g, super resolution  \cite{ledig2017photo}, image segmentation \cite{lei2020skin,xue2018segan,zhang2017deep,zhuang2016multi}, spatial transformation, denoising, etc.   Isola et. al \cite{isola2017image}  proposed pix2pix  which can be applied to general translation tasks. However, it is often difficult to obtain  paired  training data to train models of this type.

 \subsection{Unpaired Image-to-image Translation} Unpaired methods  are trained on unpaired image datasets, \textit{i.e.}, for each sample in one dataset, there are no corresponding samples in the other dataset.  Donahue et. al\cite{donahue2016adversarial} established an unsupervised generative adversarial method named BiGAN, which uses feature learning and representation to translate the data. Zhu et. al\cite{zhu2017unpaired} proposed a cycle-consistent adversarial network (Cycle-GAN), which consists of forward and backward cycle models for unsupervised image-to-image translation.  Kim et. al\cite{kim2017learning} invented DiscoGAN    to seek the relationship between source and target domains. Huang et. al\cite{huang2018multimodal} proposed MUNIT, which separates content and intensity distributions in the latent space and achieves translation by switching intensity distributions. Yan et. al\cite{yan2019domain} presented an improvement of CycleGAN, which  reduces the domain gaps  between cardiac images obtained on devices from different vendors.  \textcolor{black}{Zhang et. al \cite{zhang2019noise} proposed an approach to generate new images while preserving the main edges in the original images. Guo et.~al~\cite{guo2022alleviating} proposed SCCGAN which has a similar ability to preserve structures of source images during translation through a structure consistency constraint.} Fourier-transform-based style translation methods \cite{yang2020fda,cai2021frequency,huang2022deep} also demonstrate notable proficiency in efficiently translating images while preserving essential structural details. These techniques function by decomposing images into low-frequency and high-frequency components, with the latter capturing object structures resembling the identity. Choi et. al\cite{choi2020stargan} proposed an image-to-image translation network named StarGAN v2 based on StarGAN which can translate the image with richer textures than CycleGAN. Many other implementations of StarGAN in medical image translation have since been established. For example, Abu-Srhan et. al\cite{abu2021paired} proposed a TarGAN based on StarGAN architecture and cycle-consistency loss for multi-modal image translation, and Bashyam et. al\cite{bashyam2022deep} utilized StarGAN v2 on Brain MRI domain translations. Diffusion-based method DDIB \cite{su2022dual}, which transfers images by connecting source and target noising distributions through Schrödinger bridge. Since unpaired methods do not require paired samples, these methods are more convenient to practically apply and attract more attention for reducing domain gaps.  These unpaired methods transfer the intensity distributions between domains and  reduce the domain gaps to a certain extent. However,  existing methods mainly consider the gap between intensity distributions while failing to capture changes in structure, which is vital for medical image segmentation.

\begin{figure*}[!t]
  \includegraphics[width=1\linewidth]{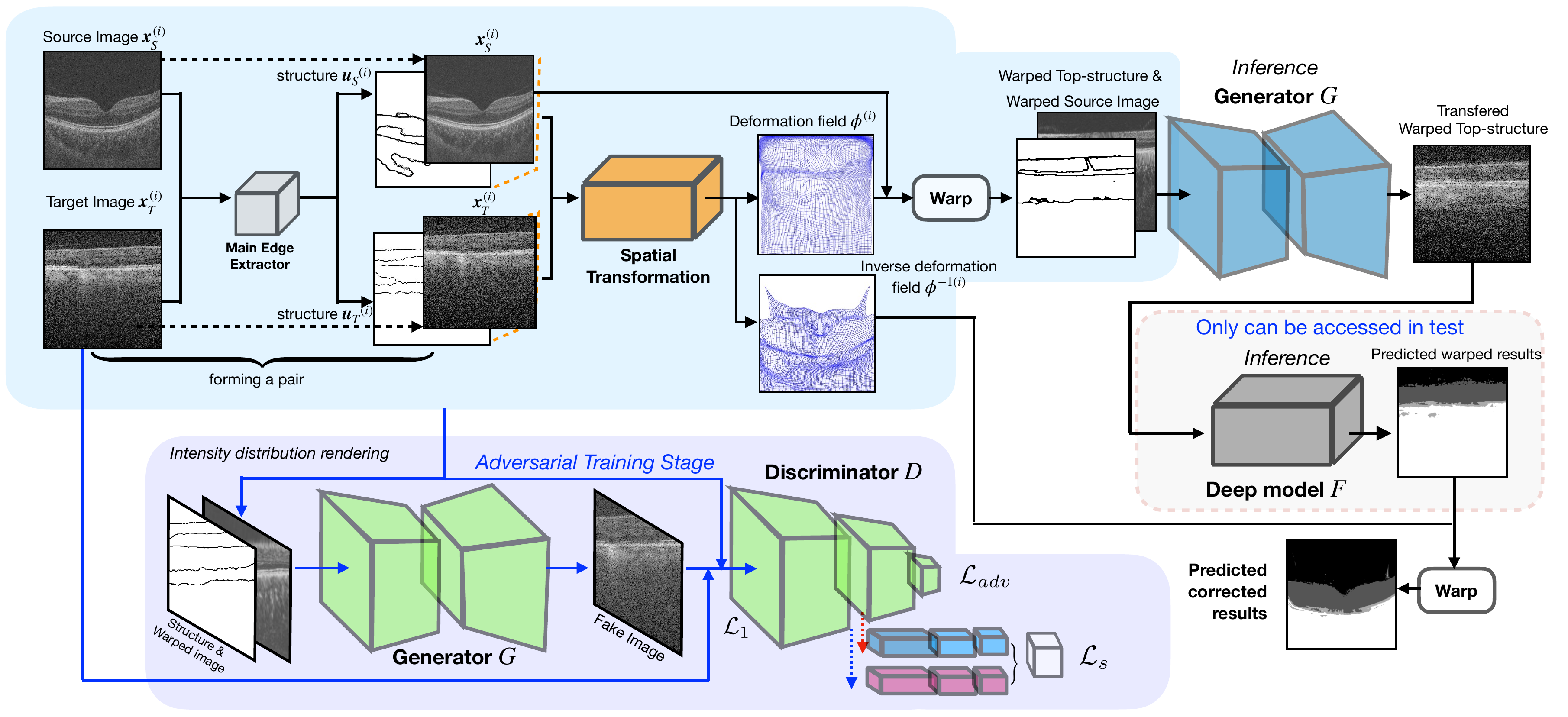}
  \caption{An illustration of the SUA: Firstly, the source and target images ($\bm x^{(i)}_{S}$ and $\bm x^{(i)}_{T}$) are processed to compute the dominant  structure of the input image. Then, the obtained dominant structures ${\bm u_{S}}^{(i)}$ and ${\bm u_{T}}^{(i)}$ are used to  compute the deformation field $\bm \phi^{(i)}$ and its inverse $\bm \phi^{-1(i)}$. The deformation field   $\bm \phi^{(i)}$ is used to warp the ${\bm u_{S}}^{(i)}$, which is further processed by the generator $G$.   The resultant   image is  fed to the trained segmentation model, whose output is   warped back by the inverse deformation field $\bm \phi^{-1(i)}$ to get the final segmentation result.}
  \label{pipline}
\end{figure*}

\color{black}
  \begin{figure*}[!t]
\center
  \includegraphics[width=1\linewidth]{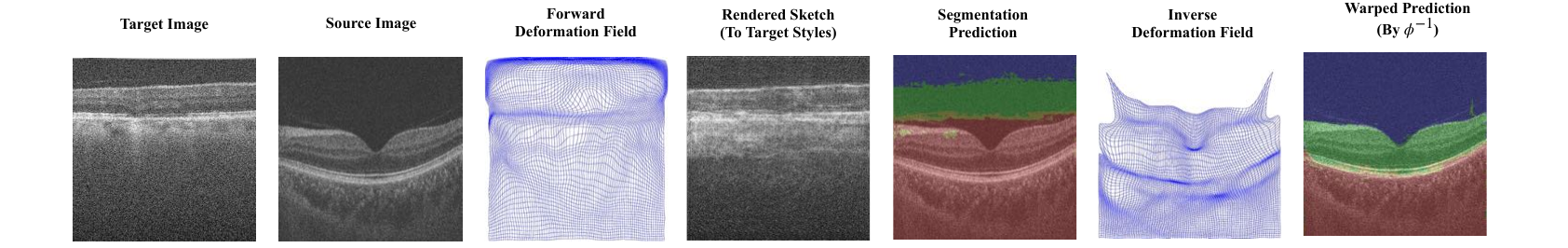}
  \caption{Intermediate results produced by the pipeline. From left to right are target image $\bm x_T$, source image $\bm x_S$, the forward deformation field $\phi$, the warped top-structure, the rendered top-structure, the prediction results, the inverse deformation field $\phi^{-1}$, and the prediction results warped back to the original structure.}
  \label{demos_joint}
\end{figure*}

\subsection{Spatial Deformation}
\color{black}
Spatial deformation estimates a mapping between a source image and a target image, and the resultant deformation can be used to warp the source image such that it has similar structure to that in the target domain. Among different approaches, diffeomorphism is an important feature which describes an invertible function that maps one differentiable deformation to another such that both the function and its inverse are differentiable \cite{beg2005computing}. Conventional iterative diffeomorphic methods such as LDDMM \cite{beg2005computing}, Dartel \cite{ashburner2007fast}, ANTs \cite{avants_ANTS}, and Demons\cite{vercauteren2009diffeomorphic} are accurate but suffer from high computational costs. Very recently,   Thorley et. al\cite{thorley2021nesterov} proposed a fast iterative method based on the Nesterov accelerated ADMM  for diffeomorphic routine, but its performance is limited for cross-domain problems. Building on the spatial transformer network \cite{jaderberg2015spatial}, the last few years have seen a boom in image spatial transformer methods based on deep learning. These include VoxelMorph\cite{dalca2018unsupervised},  SYMNet\cite{Mok_2020_CVPR}, B-spline Network \cite{qiu2021learning} and VR-Net\cite{jia2021learning}, just to name a few. To achieve diffeomorphisms, most deep learning methods use multiple squaring and scaling as a neural layer \cite{dalca2018unsupervised, Mok_2020_CVPR, qiu2021learning}. However, these methods are designed without considering differences in intensity distributions.   

\subsection{Geometry Accuracy in Synthetic Data}
\label{geometry}
\color{black}
 The assessment of geometry accuracy in synthetic data serves as a crucial evaluation metric, as highlighted by \cite{Spadea2021}. For instance, the Dice similarity coefficient is commonly computed to evaluate the accuracy of representing specific tissue classes or structures such as bones, fat, muscle, air, and the overall body. It has been observed that registration errors in synthetic data can introduce blurring artifacts in high-contrast regions, leading to inaccuracies in treatment localization, as noted by \cite{osti_22250645}. In certain applications, such as subsequent segmentation tasks, achieving structure correspondence becomes essential. Jiang et al. \cite{Jiang2019} investigated the use of MRI-to-CT translation to enhance the robustness of segmentation, while Kieselmann et al. \cite{Kieselmann2021} generated synthetic MRI data from CT scans to train segmentation networks for reliable auto-segmentation algorithms of organs-at-risk and radiation targets. Additionally, Zhang et al. \cite{zhang2019noise} explored the adaptation segmentation task for medical images, emphasizing the importance of geometry accuracy in medical image translation when applying a trained segmentation model to fit images from a different domain by translating them to the source domain. In Section \ref{results} of our work, we have demonstrated the efficiency of our proposed method in this segmentation task.

\section{Method}
\label{method}
In this paper, we propose an image translation approach such that an input image can be converted to an output image while the domain gaps in both intensity distribution and structure can be minimized. We define the test (source) data to be translated as $X_S$ and the set of training (target) data with labelled segmentation ground truth  as $X_T$. Our goal is to learn a mapping from $X_S$ to $X_T$, such that the element $\bm x_S \sim P_{X_S}$ will be translated to $\bm x_T \sim P_{X_T}$ which has the same underlying structure and intensity distribution as in $X_T$. Since the objective of the intensity distribution translation is to overcome the domain gap issue for subsequent analysis tasks such as segmentation, the content of the images or the underlying clean part of the images is what really matters and shall be kept unchanged. However, arbitrarily maintaining the edges would lead to the propagation of the structural gap. In this paper,   we first obtain the main structure  $\bm u$ via a preprocessing step. Then, a spatial transformation is used to get the forward and inverse deformation fields ($\phi$ and $\phi^{-1}$) between the input and target images. After that, we utilize  $\phi$ to warp structure masks which are formed by filling the structural images to get  $\bm u(\phi)$ which we re-obtain the edges from to get the warped structure and feed into the structure-to-image rendering generator $G$.  Finally, the outputs $G(\bm u(\phi),\bm x_S(\phi))$ of $G$ are expected to have  similar structures and intensities to images in the target domain. The overall process is  shown in Fig.~\ref{pipline}. Since the structures are deformed in the spatial transformation block, we use the inverse deformation field $\phi^{-1}$ to warp the segmentation outputs back to original shape as the final segmentation output. An illustration of the intermediate results  is given  in Fig.~\ref{demos_joint} for a better understanding of the proposed method.

\subsection{Pixel Clustering Map}\label{Clustering Map}
\begin{figure}[!t]
  \includegraphics[width = 1 \linewidth]{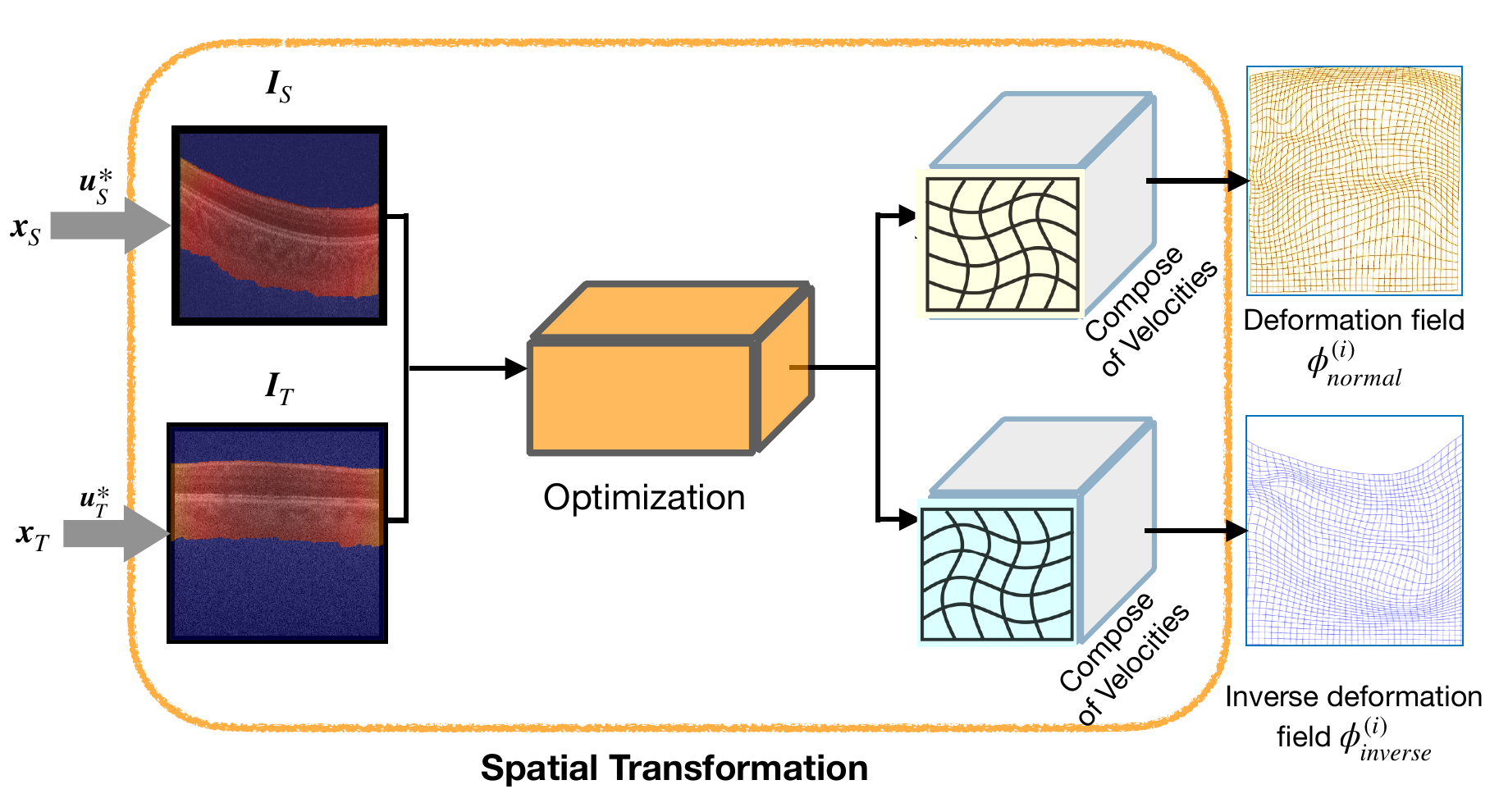}
  \caption{Illustration of the spatial transformation procedure.}
  \label{registration}
\end{figure}

We first compute a basic pixel clustering map $\bm u^*$ for input image $\bm x$ using the joint image reconstruction and pixel clustering Potts model \cite{storath2015joint}. Then its edge sketch $\bm u$ is obtained. Meanwhile, we combine the clustering map into binary masks. Then these masks are given a Gaussian gradient and multiplied by their corresponding source images $\bm x_{S}$ and $\bm x_{T}$ to get the composed structure images $\bm I_{S}$ and $\bm I_{T}$, which is shown in Fig \ref{registration}. 

\subsection{Spatial Transformation Block}
In this section, we establish a spatial transformation block which obtains a deformation to reduce the domain shift between dominant structures in the datasets. We note that most related methods only include a forward deformation estimation strategy. However, an inverse deformation is crucial for our application in image segmentation. If we perform an inverse computing by switching source and target, the obtained inverse deformation will not match the forward deformation well, thus, we propose a spatial transformation block which learns the forward and inverse deformation fields simultaneously.

\color{black}

After conducting the step mentioned in Section \ref{Clustering Map}, a pair of the composed structure images $\bm I_S$ and $\bm I_T$ (the pair with the maximum SSIM score) are obtained. Here, we propose an invertible spatial transformation to align them. Since we do so by multiplying Gaussian masks with the original images, this can be considered performing a transformation on the masks rather than a cross-domain transformation.    Computing a diffeomorphic deformation can be treated as modelling a dynamical system \cite{beg2005computing}, given by an ordinary differential equation (ODE): $\partial{\bf{\phi}}/\partial t  = {\bf{v}}_t({\bf{\phi}}_t)$, where ${\bf{\phi}}_0 = {\rm{Id}}$ is the identity transformation and ${\bf{v}}_t$ indicates the velocity field at time $t$ ($\in [0,1]$).  To solve the ODE, we use Euler integration, in which the forward deformation field $\phi$ is calculated as the compositions of a series of small deformations, defined as
\begin{equation} \label{eq:ode}
    \phi=\left({\rm{Id}}+\frac{{\bf{v}}_{t_{N-1}}}{N}\right)
    \circ \cdots \circ \left({\rm{Id}}+\frac{{\bf{v}}_{t_{1}}}{N}\right)\circ \left({\rm{Id}}+\frac{{\bf{v}}_{0}}{N}\right).
\end{equation}
The backward deformation can be computed reversely as 
\begin{equation} \label{eq:ode}
    \phi^{-1}= \left({\rm{Id}}-\frac{{\bf{v}}_{0}}{N}\right)
    \circ \left({\rm{Id}}-\frac{{\bf{v}}_{t_{1}}}{N}\right)
    \circ \cdots 
    \circ \left({\rm{Id}}-\frac{{\bf{v}}_{t_{N-1}}}{N}\right).
\end{equation}

\begin{figure}[!t]
\center
  \includegraphics[width = 1 \linewidth]{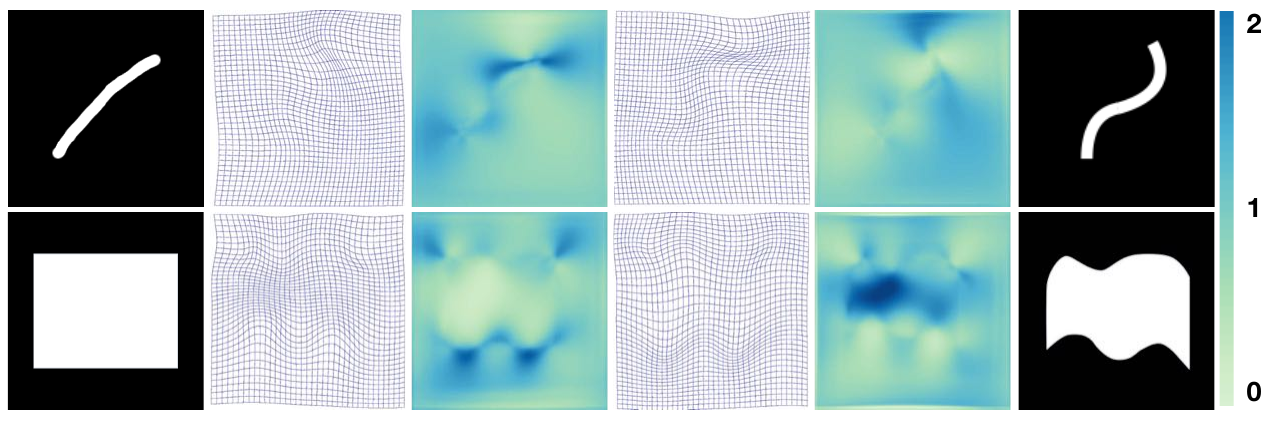}
  \caption{Visualisations of diffeomorphic forward and backward (inverse) deformations. The first and the last columns show the source and target image, respectively. The second and the fourth columns  show the forward and backward deformation fields, respectively. Finally, the third and the fifth columns show the Jacobian determinants of corresponding forward and backward deformations, respectively.  }
  \label{ADMM_demo}
\end{figure}

In the above equations, if the velocity fields ${\bf{v}}_{t_{i}}, \forall i\in \{0, ..., N-1\}$ are sufficiently small whilst satisfying some smoothness constraints, the resulting composition is a diffeomorphic deformation. In addition, note that the composition between $\phi$ and $\phi^{-1}$ will give an approximate identity grid and one can use $\phi^{-1}$ to warp an image back. 

To compute the velocity fields whilst satisfying these diffeomorphic constraints, we use the following model 
\begin{equation} \label{eq:OF}
\min_{\bf{v}} \left \{ \frac{1}{2} \|\rho({\bf{v}}) \|^2 + \frac{\lambda}{2} \|\nabla^n {\bf{v}} \|^2 \right \},
\end{equation}
where $\rho({\bf{v}})= \langle \nabla \bm I_S, {\bf{v}} \rangle + \bm I_S - \bm I_T $. $\nabla^n$ denotes the $n^{\mathrm{th}}$ order gradient, and here we use $n=3$. For a scalar-valued function $f(x,y,z)$ in the continuous setting, $\nabla^n f =  \arraycolsep=10.5pt\def\arraystretch{0.1} 
[ (\begin{array}{c}
n\\
k_1,k_2,k_3
\end{array})\frac{\partial^2 f}{\partial x^{k_1}\partial y^{k_2} \partial z^{k_3} }]^{\rm{T}}$, where $k_i=0,...,n$, $i \in \{1,2,3\}$ and $k_1+k_2+k_3=n$. Moreover, $\arraycolsep=10.5pt\def\arraystretch{0.1} (\begin{array}{c}
n\\
k_1,k_2,k_3
\end{array})$ are known as the multinomial coefficient which is computed by $\frac{n!}{k_1!k_2!k_3!}$. To solve the model effectively, we use a multiscale ADMM algorithm developed in \cite{thorley2021nesterov}. An illustration of forward and inverse deformations is shown in Fig.~\ref{ADMM_demo}.

\subsection{Intensity Distribution Rendering}

\begin{figure}[!t]
  \includegraphics[width = 1\linewidth]{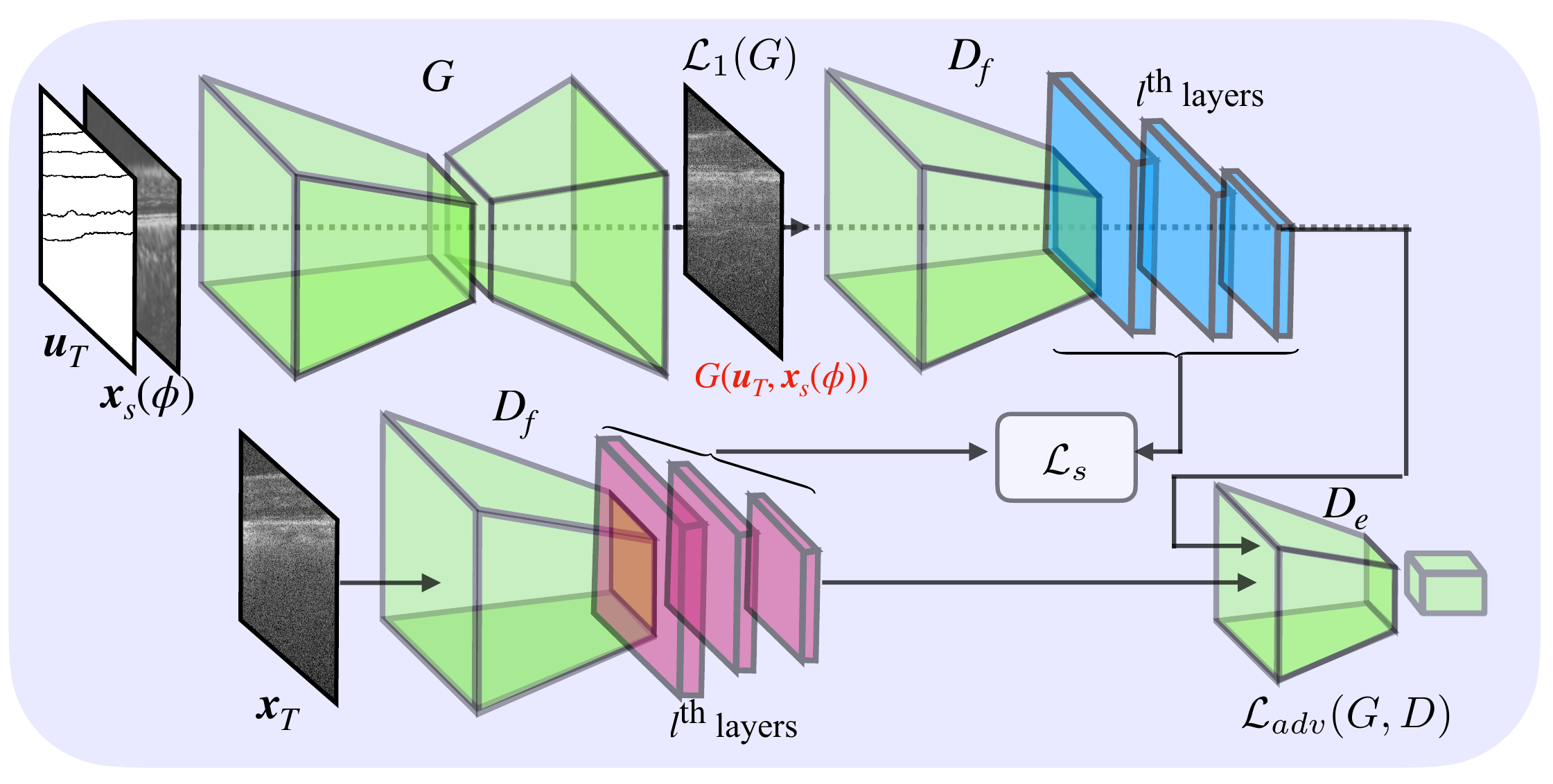}
  \caption{Architecture of the intensity distribution rendering network, where $D_f$ and $D_e$ denotes the start and the end parts of the discriminator $D$.}
  \label{rendering}
\end{figure}

An intensity distribution rendering network is used to render the warped structures with the targeted intensity distribution.   As shown in Fig.~\ref{rendering}, it is a paired image-to-image translation network, including a "U-Net-like"\cite{ronneberger2015u} structure for translation mapping, a discriminator for adversarial training,   and a feature alignment mechanism for loss computation.  Mathematically, the losses $\mathcal{L}_{adv}$, $\mathcal{L}_1$ and $\mathcal{L}_s$ corresponding to the three components are denoted as:
\begin{align}
\label{eq4}
		\mathcal{L}_{adv}&=\mathbb{E}_{\bm u_T \sim U_T}[\log (1-D(G(\bm u_T,\bm x_S( \phi)),\bm u_T))]\\
	&+\mathbb{E}_{(\bm x_T,\bm u_T)\sim (X_T,U_T)}[\log D(\bm x_T,\bm u_T)], \notag
\end{align}
where $\bm u_T\sim U_T$ represents the targeted structure, $(\bm x_T,\bm u_T)\sim(X_T,U_T)$ represent the targeted image and structure pair and $D$ represents the discriminator of the mapping $G$, and $\phi$ represents the deformation between $\bm x_T$ and its closest $\bm x_S$ (maximum SSIM).

\begin{equation}
    \mathcal{L}_1 = \mathbb{E}_{(\bm x_T,\bm u_T)\sim (X_T,U_T)}\|G(\bm u_T,\bm x_S( \phi))-\bm x_T\|_1,\label{eq5}
\end{equation}
where $\|\cdot\|_1$ denotes the $\ell_1$-norm. Then feature correlations are given by Gram matrix $Gr$, where $Gr(\bm x_T)_{ij}^{\left( l\right)}$ correspond to the Gram matrix of $l^{th}$ layer of $D$ with input $\bm x_T$ and $\bm u_T$. It is computed as follows:
\begin{equation}
    G r(\bm x_T)_{ij}^{\left( l\right)}=\operatorname{vec}\left[\mathcal{F}(\bm x_T)_{i}^{\left(l\right)}\right] ^T \operatorname{vec}\left[\mathcal{F}(\bm x_T)_{j}^{\left(l\right)}\right],
\end{equation}
where $\mathcal{F}(\bm x_T)_{i}^{\left(l\right)}$ represents the $l^{th}$ layer of $D$'s feature maps, the subscripts $i$ and $j$ denote the $i^{th}$ and $j^{th}$ channel, and $\text{vec}(\cdot)$ denotes the vectorization operation.
\begin{equation}
    \resizebox{0.9\hsize}{!}{$\mathcal{L}_{s}=\mathbb{E}_{(\bm x_T,\bm u_T)\sim (X_T,U_T)} \sum_{l=1}^{3}\|G r(G(\bm{u}_T,\bm x_S( \phi)))^{ (l)}-G r(\bm{x}_T)^{( l)}\|_{F}$}
    \label{eq6}
\end{equation}
The full objective of the intensity distribution rendering network is described as the following equation.
\begin{equation}
	G^{*},D^{*}=\arg \min_{G} \max_{D}~[\mathcal{L}_{adv}+\lambda_1\mathcal{L}_{1}+\lambda_2\mathcal{L}_{s}],
	\label{eq7}
\end{equation}
where $\lambda_1$ and $\lambda_2$ are hyper-parameters that balance different losses.  The optimization procedures of the whole model are summarized in Algorithm~\ref{Algorithom}.

		
		
		
			
				


			
		

\begin{algorithm}[t]
   \caption{Optimization Procedures} 
   \textbf{Input:} $\bm X_S$ and  $\bm X_T$.\\
   \textbf{Require:} $G$ and $D$. Two Adam optimizers $A_1$ 
    and $A_2$.  \\
   \For{epoch$<N$}  { 
         \If{$epoch=N-1$}
      {
        \For{  $\bm x_{S}^{(j)}\in \bm S$}
        {		 Obtain $\bm u_{S}$ and  $\bm u^*_{S}$; \\
    		Calculate the SSIM of $\bm I_{S}^{(i)}$ and $\bm I_{T}^{(j)}$;
        }
        Solve deformations $\phi$ and $\phi^{-1}$ for indices of the pair that has the maximum SSIM via (\ref{eq:OF});
       }
    
	\For{  $\bm x_{S}^{(i)} \subset \bm X_{S}$ }
         {
		\For{  $\bm x_{T}^{(j)}\subset \bm  X_{T}$ }   
            {   Obtain $\bm u_{T}^{(i)}$;
			  Calculate $\mathcal{L}_{adv}(G,D)$, $\mathcal{L}_{s}(G)$ and $\mathcal{L}_{1}(G)$;
				$D\leftarrow A_1(D,[\bm x_{T}^{(j)},\bm u_{T}^{(i)}],\mathcal{L}_{adv})$;
				$G\leftarrow A_2(G, D,[\bm x_{T}^{(j)},\bm u_{T}^{(i)}],[\mathcal{L}_{adv},\mathcal{L}_{1},\mathcal{L}_{s}])$;
            }
	}
 }

\textbf{Output:} $G$, $\phi$, $\phi^{-1}$. 
	\label{Algorithom}
\end{algorithm}

\subsection{Implementation}
\noindent \textbf{Network Architecture:} 
The generator in our networks includes 15 residual 2D convolution blocks. Specifically, the first 8 blocks are encoder blocks, where each block contains a 4$\times$4 convolution layer sequentially followed by a ReLU activation function, an instance normalization layer and a dropout layer; the rest are decoder blocks, where each block contains a 4$\times$4 transposed convolution layer sequentially followed by a ReLU function, an instance normalization layer and a dropout layer, respectively. The discriminator in the proposed model contains four convolution blocks and a fully connected layer, where these convolution blocks include a  4$\times$4 convolution layer (padding equals to 1), an instance normalization layer and a Leaky-ReLU function. The core code will be released after the acceptance of our work.

\noindent \textbf{Training Details:} In intensity distribution rendering, we set the balancing hyper-parameter $\lambda_1=1$ and $\lambda_2=100$ in equation (\ref{eq7}) for all the  experiments. We apply the Adam \cite{Adam} optimizer with a learning rate of $0.0002$, which decays to zero following a linear principle from the $100^{th}$ epoch to the $200^{th}$ epoch in all experiments. Moreover, the padding pixel number is $8$, the tolerance ratio is $0.001$, the balance hyper-parameter is $5$ and the max-iteration is $50$ in the spatial transformer of all experiments. Additionally, we set the $\gamma = 0.35$ for the OCT and the cardiac experiments, and $\gamma=0.55$ for the MRI and CT experiments in the Potts model.

\section{Experimental Results}
\label{results}
We conduct a comprehensive evaluation of the Structure-Unbiased Adversarial (SUA) network across multiple domain adaptation tasks in medical image segmentation. In retinal Optical Coherence Tomography (OCT), we utilize the SINA and ATLANTIS datasets to address structural and intensity distribution differences. Similarly, in MRI-to-CT domain transfer, we leverage publicly available datasets to demonstrate the method's applicability to multi-modal imaging. Additionally, we assess the adaptability of SUA to pathological conditions by transferring images from diseased (ACDC) to healthy (UKBB) subjects in cardiac MRI. Through extensive experiments, we evaluate both structure and intensity translation, employing various evaluation metrics and comparisons with state-of-the-art methods. These experiments collectively showcase the effectiveness and versatility of our SUA in addressing domain gaps in medical image segmentation tasks.
\subsection{Domain Translation on Retinal OCT} \label{octexp}
	\subsubsection{Datasets}
	 We apply SUA on retinal OCT to transfer the 
	shape characteristics and intensity distribution of one OCT dataset to another.  The SINA \footnote{\url{https://people.duke.edu/~sf59/Chiu\_IOVS\_2011\_dataset.htm}} and ATLANTIS datasets are used.
The   SINA  dataset contains 220 B-scans from 20 volumes of eyes with drusen and geographic atrophy, collected using a spectral domain-OCT imaging system from Bioptigen. Three  boundaries  have been manually annotated, including boundary 1: internal limiting membrane (ILM); boundary 2: between the outer segments and the retinal pigment epithelium (OS/RPE); and boundary 3: between Bruchs membrane and the choroid (BM/Choroid). We use this dataset as the  target $T$.
	The  ATLANTIS dataset is a local dataset, containing 176 B-scans,    collected using a swept-source OCT machine. The same three  boundaries as SINA are annotated and used in this study. 
  Since the two datasets are collected from different machines under different protocols, there exist gaps in both structure and intensity distributions, which makes the models trained from one dataset perform poorly on the other.  

  	\begin{table*}[!t]
	\normalsize
	\setlength{\tabcolsep}{0.75 mm}
	\caption{Distribution and Segmentation Evaluation of Retinal OCT}
    \vspace{-10pt}
	\begin{center}
	\begin{tabular}{lp{1.2cm}p{1cm}p{1.9cm}p{1.9cm}p{1.9cm}p{1.9cm}p{1.9cm}p{1.9cm}}
	    \toprule[1pt]
		\multirow{2}{*}{Methods} & \multicolumn{2}{c}{Distribution} & \multicolumn{6}{c}{Segmentation} \\ 
        \cmidrule(lr){2-3} \cmidrule(lr){4-9}
		& $D_{Bhat}$ $\downarrow$ & $Corr$   $\uparrow$  &$Acc \uparrow$ & $Dice \uparrow$ & $mIoU \uparrow$ &  $Sen \uparrow$ & $Spe \uparrow$ & $FDR \downarrow$  \\
		\cmidrule(lr){1-3} \cmidrule(lr){4-9}
		Without Translation &
		
0.391 &0.587 &
0.854 (0.028) &
0.538 (0.060) &
0.520 (0.057) &
0.586 (0.056) &
0.891 (0.023) &
0.384 (0.027) \\

		VoxelMorph \cite{dalca2018unsupervised} &
0.340 &0.369 &		
0.760 (0.013) &
0.324 (0.016) &
0.235 (0.015) &
0.500 (0.015) &
0.806 (0.006) &
0.533 (0.115)
\\
		VoxelMorph+MI \cite{dalca2018unsupervised} &
0.367 &0.586 &	
 0.785 (0.008) &
 0.501 (0.026)&
 0.281 (0.018)&
0.561 (0.036) &
0.833 (0.005) &
0.558 (0.022) 
\\
		VR-Net \cite{jia2021learning} &
		0.237&0.677&
0.761 (0.015) &
0.333 (0.036) &
0.240 (0.027) &
0.515 (0.031) &
0.812 (0.009) &
0.551 (0.058)    \\
		CycleGAN \cite{zhu2017unpaired}  &0.139& 0.810 &
0.875 (0.024) &
0.624 (0.058) &
0.597 (0.059) &
0.656 (0.065) &
0.909 (0.017) &
0.363 (0.052)
\\
		MUNIT \cite{huang2018multimodal} &
\textbf{0.071}&\textbf{0.959} &
0.811 (0.021) &
0.522 (0.033) &
0.323 (0.032) &
0.539 (0.026) &
0.867 (0.014) &
0.529 (0.021)
\\
		DualGAN \cite{yi2017dualgan} &
		0.313& 0.597&
0.839 (0.014) &		
0.500 (0.024) &
0.390 (0.021) &
0.574 (0.018) &
0.881 (0.010) &
0.394 (0.012) 
\\
		NAGAN \cite{zhang2019noise}&
0.177& 0.750 &
0.955 (0.025) &
0.748 (0.032) &
0.681 (0.053) &
0.771 (0.034) &
0.969 (0.021) &
0.250 (0.021) 
\\
		StarGAN v2 \cite{choi2020stargan} &
		0.244& 0.724&
0.754 (0.034) &		
0.312 (0.044) &
0.231 (0.044) &
0.335 (0.044) &
0.823 (0.027) &
0.639 (0.050) 
\\
\textcolor{black}{SCCGAN} \cite{guo2022alleviating}  &
0.354	&0.647 &
0.945 (0.007)&
0.749 (0.045)&
0.621 (0.048)&
\textbf{0.847} (0.052)&
0.955 (0.008)&
 0.298 (0.032)
\\

		DDIB \cite{su2022dual} &
 0.355&0.598&		
 0.881 (0.052)&
 0.560 (0.098)&
 0.569 (0.104)&
 0.567 (0.090)&
 0.918 (0.036)&
 0.527 (0.085)\\
  
\cmidrule(lr){1-3} \cmidrule(lr){4-9}
		\textbf{Ours} & 
0.102& 0.874 &
\textbf{0.980} (0.008) &
\textbf{0.816} (0.007) &
\textbf{0.763} (0.087)&
0.804 (0.072) &
\textbf{0.985} (0.006) &
\textbf{0.165} (0.067) 
\\
		\bottomrule[1pt]           
	\end{tabular}
	\end{center}
	\label{tableOCT}
\end{table*}
\begin{figure*}[h!]
    \centering
    \includegraphics[width = \textwidth]{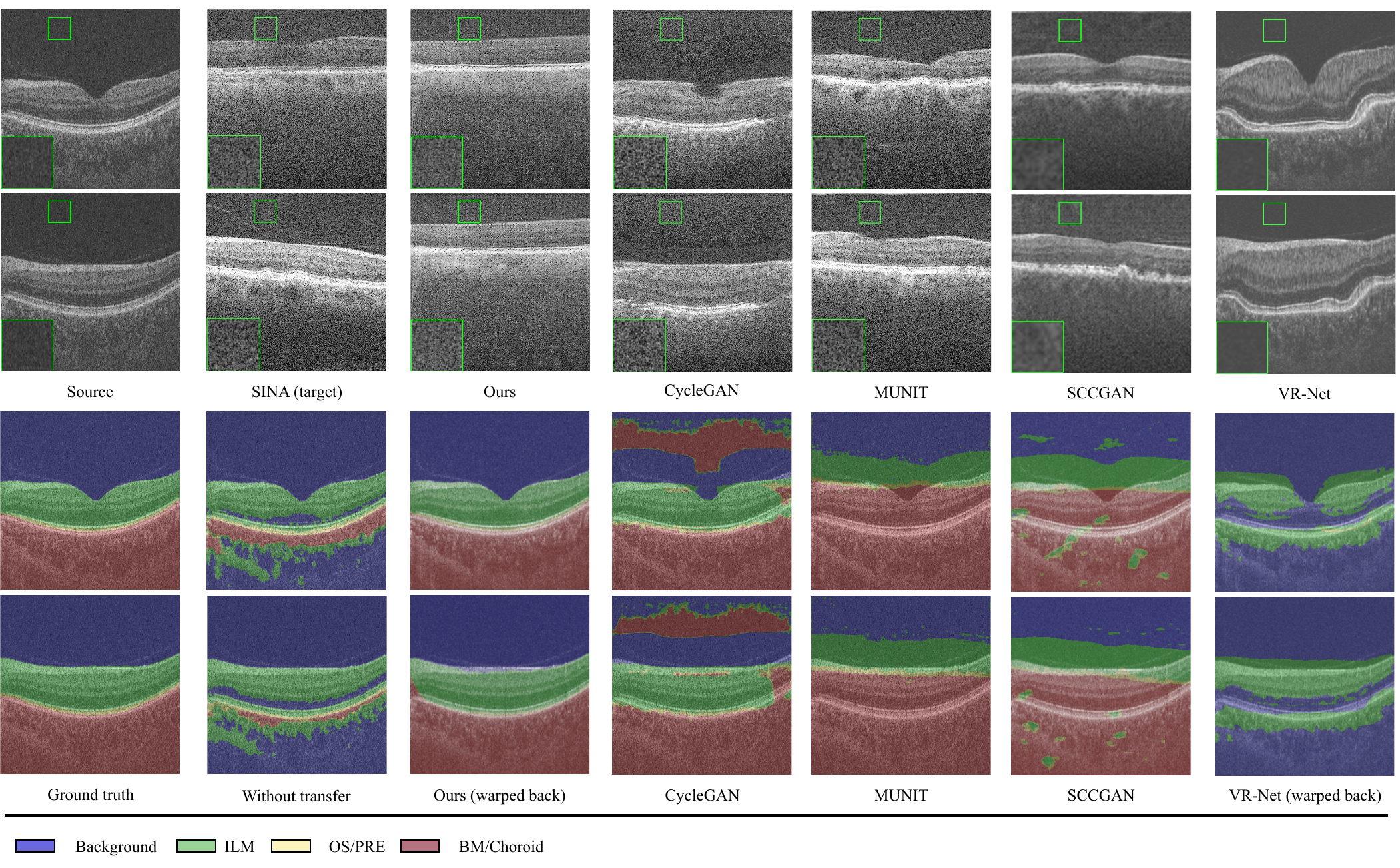}
    \vspace{-10pt}
    \caption{Results of translated images from ATLANTIS to SINA. The first and second rows show translated images of two input images by different methods.  The third and fourth rows show the corresponding segmentation results.}
    \label{fig:OCT_trans}
\end{figure*}
	
	
	\subsubsection{Comparison with prior arts}
	The performance of the image translation is evaluated in terms of both structure and intensity distribution.

	\textbf{{Evaluation on Structure Translation}}: The performance evaluation of the structure transfer is challenging. Ideally, it would be evaluated by comparing the translated image with the corresponding image from the other machine. However, it is not practical in image to identify the exact same region of the object using different machines. Since the main objective of the translation is to improve subsequent analysis, we evaluate its performance using the segmentation results indirectly.

A segmentation network based on the U-Net architecture is first trained using the SINA dataset (training/target dataset). The proposed generative model is trained to translate ATLANTIS images such that the   structure characteristics and intensity distributions of the translated images are similar to those in SINA. The trained U-Net segmentation model is then applied on the translated data (test/source dataset)  to detect the three boundaries, which are subsequently used to evaluate the performance of the image-to-image translation model. 

	\begin{figure*}[!t]
    \centering
    \includegraphics[width=\textwidth]{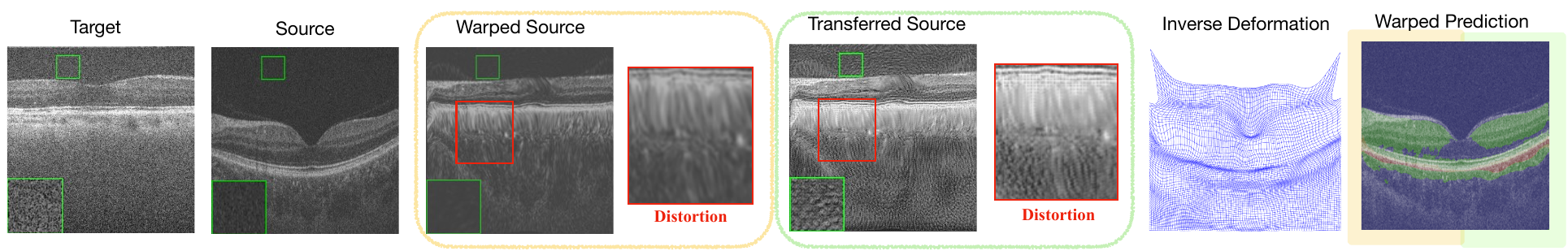}
    \vspace{-10pt}
    \caption{Illustrations of results and deformations in the ablation study. It shows the reasons (distortions caused by deformations) why a straight forward combination of spatial transformation model and structure-preserving GAN will not work well.}
    \label{fig:ablation}
\end{figure*}

The mean intersection-over-union ($mIoU$) \cite{khoreva2017simple}, the mean Sørensen–Dice coefficient ($Dice$), the accuracy ($Acc$) , the sensitivity ($Sen$), the specificity ($Spe$), and the false discovery rate ($FDR$) are used as evaluation metrics.  

To evaluate the effectiveness of the proposed SUA network,   we compare it with state-of-the-art   translation methods including CycleGAN \cite{zhu2017unpaired}, MUNIT \cite{huang2018multimodal}, DualGAN \cite{yi2017dualgan}, StarGAN v2 \cite{choi2020stargan} NAGAN\cite{zhang2019noise}, \textcolor{black}{SCCGAN \cite{guo2022alleviating}} and DDIB \cite{su2022dual}  as well as the   spatial transformation methods  VoxelMorph \cite{dalca2018unsupervised}, VoxelMorph with mutual information loss (VoxelMorph+MI) and VR-Net \cite{jia2021learning}. For all SOTA methods and the test tool U-net, we use the same hyper-parameters described in the open-sourced codes or the papers from the original authors. In the experiments, we train these GANs in a similar way to learn the translation from ATLANTIS to SINA. The transferred ATLANTIS is then fed into the U-Net segmentation network to segment the three boundaries for comparison with the manual ground truth. For the spatial transformation methods, the   deformation is computed to warp the image for segmentation. The output of the segmentation is warped back by the inverse deformation for comparison. TABLE  \ref{tableOCT} shows the comparison between the proposed method and other methods, where values are shown in the form of mean(std). As shown from the results, the proposed SUA outperforms the state-of-the-art performance on all metrics.

	Fig.~\ref{fig:OCT_trans} shows some sample results for visual comparison. As can be observed, 
 Cycle-GAN indiscriminately transfers both intensity distribution and structure, resulting in a SINA-shaped segmentation input and prediction. Since the prediction is in a different domain than the input, we cannot forcibly apply SINA-to-ATLANTIS of Cycle-GAN and there is no displacement information as we obtained in our method. Therefore, the raw SINA-shaped prediction shows very low performance. This phenomenon is  more obvious on MUNIT and $\text{StarGAN v2}$ which transfer the structure even more effectively. Respectively, our method achieves a gain of 0.251 on $mIoU$ and 0.189 on $Dice$  compared to Cycle-GAN, and a gain of 0.525 on $mIoU$ and 0.391 on $Dice$  compared to MUNIT. \textcolor{black}{Similarly to the NAGAN, SCCGAN has the ability to preserve structures through a structure consistency constraint based on squared-loss Mutual Information.}
	There are two reasons why DualGAN does not achieve desired segmentation performance. Firstly, they cannot differentiate the difference between structure and intensity distribution. Secondly, they suffer from the domain gap in shape which we tackled using the spatial transformation block. This happens to NAGAN \textcolor{black}{and SCCGAN} as well, and our method achieves a gain of 0.358 and 0.067 on $mIoU$, 0.313 and 0.065 on $Dice$  compared to DualGAN and NAGAN respectively. The superior performance of our proposed method over MUNIT stems from a particular drawback in MUNIT. Specifically, MUNIT translates the style by switching the style features in the latent space and doesn't have spatial movement predictions, which struggles to track changes in structure and shape during the translation process. As a result, the altered structure and shape do not align well with the original image, leading to diminished segmentation performance. Similar to MUNIT, DDIB struggles to track changes in structure and shape during the translation process, which decrease the segmentation results by lacking correspondence. In comparison with NAGAN, the SCCGAN is designed to keep the structure and shape unchanged during the translation, which maintains the structure gap to the target dataset. Therefore, it does face challenges in reducing the domain gaps related to the structure. These remaining structural domain gaps subsequently lead to reduced segmentation performance. 
	Moreover, spatial transformation methods, \emph{i.e,} Voxelmorph and VR-Net, do not achieve acceptable results because they ignore the difference in texture and intensity distribution.

	\textbf{{Evaluation on Intensity Distribution Translation}}:
	We also evaluate the performance of   intensity distribution translation from the  {SINA}  to {ATLANTIS}.  Two metrics Bhattacharyya distance  $D_{Bhat}$ \cite{bhattacharyya} and correlation  $Corr$ \cite{Gonzalez:2006:DIP:1076432} are computed to evaluate the performance of intensity distribution transfer referring to \cite{zhang2019noise}.  To calculate the two metrics, we calculate the intensity histograms of both transferred images and target images:
	\begin{equation}
	\small
	Corr(H_1,H_2)=\frac{\sum_{I}(H_1(I)- \bar{H_1})(H_2(I)- \bar{H_2})}{\sqrt{\sum_{I}(H_1(I)- \bar{H_1})^2 \sum_{I}(H_2(I)- \bar{H_2})^2}},
	\end{equation}
	\begin{equation}
	\small
	D_{Bhat}(H_1,H_2) = \sqrt{1-\frac{1}{\sqrt{\bar{H_1}\bar{H_2}N^2}}\sum_{I}\sqrt{H_1(I)H_2(I)}},
	\end{equation}
	where $H_{1}$ and $H_{2}$ denote the normalized histograms.  $I$ denotes background regions, $\bar H_k=\frac{1}{N}\sum_{J}H_k(J)$ represents the mean value of histogram $H_k,~ k=1, 2,$ and $N$ denotes the total number of histogram bins.

\begin{figure}[h]
    \centering
    \includegraphics[width=\linewidth]{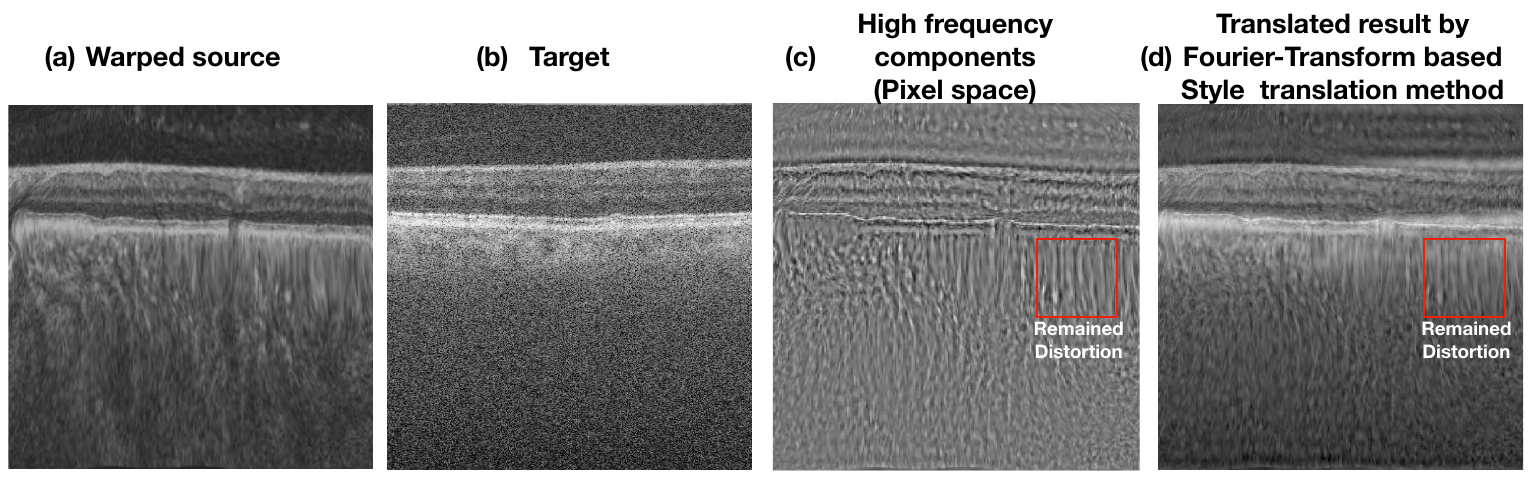}
    \vspace{-10pt}
    \caption{Illustrating distortion artifacts caused by the Fourier-transform-based style translation. (a) shows a warped source image as input to the intensity translation module; (b) is the target style we want (a) to transfer to; (c) demonstrates high-frequency sensitivity to deformations, whereas (d) showcases the retention of distortions, as highlighted in the red boxes.}
    \label{fig:Fourier}
\end{figure}
\begin{table}[!h]
	\setlength{\tabcolsep}{2 mm}
	\center
	\normalsize
	\caption{Results of Ablation Study}
		
	\begin{tabular}{l||c|c|c}
			\toprule[1.5pt]
		Method  
		& $mIoU \uparrow$ & $Dice \uparrow$   & $FDR \downarrow$ \\\hline
		w/o Translation &0.528 & 0.547&0.381\\
		DiffR & 0.367& 0.504&0.391\\
		DiffR + $u^*$  &0.316 & 0.545&0.519\\
		DiffR + GAN   & 0.305 & 0.508 & 0.530\\
		DiffR + GAN + $u^*$  & 0.326 & 0.538 & 0.525 \\
  Ours w/o SSIM \& $\bm x_s(\phi)$ &  0.613&  0.702&  0.251  \\
  Ours w/o  $\bm x_s(\phi)$ &  0.748 & 0.813 & 0.202 \\
  Ours + Fourier &0.644 &0.732  &0.259\\
  \hline
		Ours  &\textbf{0.763} &\textbf{0.816} &\textbf{0.165}\\
		\bottomrule[1pt]         
	\end{tabular} \label{table:ablation}
\end{table}

	\begin{table*}[!h]
	\normalsize
	\setlength{\tabcolsep}{0.75 mm}
	\caption{Distribution and Segmentation Evaluation of Chest MRI to CT}
	\begin{center}		
	\begin{tabular}{lp{1.2cm}p{1cm}p{1.9cm}p{1.9cm}p{1.9cm}p{1.9cm}p{1.9cm}p{1.9cm}}
	    \toprule[1pt]
		\multirow{2}{*}{Methods} & \multicolumn{2}{c}{Distribution} & \multicolumn{6}{c}{Segmentation} \\ 
        \cmidrule(lr){2-3} \cmidrule(lr){4-9}
		& $D_{Bhat}$ $\downarrow$ & $Corr$   $\uparrow$ &$Acc \uparrow$ & $Dice \uparrow$ & $mIoU \uparrow$ &  $Sen \uparrow$ & $Spe \uparrow$ & $FDR \downarrow$  \\
		\cmidrule(lr){1-3} \cmidrule(lr){4-9}
		Without Translation &
0.592 &0.043&		
0.832 (0.003) &
0.337 (0.008) &
0.253 (0.006) &
0.325 (0.012) &
0.796 (0.004) &
0.635 (0.012) 
\\

		VoxelMorph \cite{dalca2018unsupervised}&
0.371 &0.125 &
0.859 (0.003) &
0.506 (0.009) &
0.314 (0.005) &
0.520 (0.012) &
0.834 (0.003) &
0.582 (0.013)  \\

		VoxelMorph+MI \cite{dalca2018unsupervised} &
 0.328& 0.132&
0.827 (0.006) &
0.344 (0.016) &
0.257 (0.011) &
0.348 (0.025) &
0.803 (0.007) &
0.645 (0.018) 
\\

		VR-Net \cite{jia2021learning}&0.346&0.135&
0.870 (0.004) &
0.541 (0.015) &
0.336 (0.012) &
0.507 (0.015) &0.849 (0.005) &
0.557 (0.017) 
\\

		CycleGAN \cite{zhu2017unpaired}  &
0.300& 0.314 &
0.930 (0.014) &
0.683 (0.076) &
0.549 (0.073) &
0.781 (0.092) &
0.938 (0.024) &
0.343 (0.046) \\
		MUNIT \cite{huang2018multimodal} &
\textbf{0.286} &\textbf{0.507} &
0.924 (0.015) &
0.651 (0.086) &
0.517 (0.080) &
0.744 (0.104) &
0.929 (0.026) &
0.364 (0.055) 
\\

		DualGAN \cite{yi2017dualgan}&
0.386& 0.102 &
0.818 (0.004) &
0.348 (0.007) &
0.260 (0.010) &
0.358 (0.009) &
0.816 (0.012) &
0.635 (0.006) 
\\
		NAGAN \cite{zhang2019noise}&
0.320&0.059 &
0.841 (0.037) &
0.557 (0.066) &
0.532 (0.067) &
0.608 (0.046) &
0.885 (0.027) &
0.389 (0.035) 
\\
		StarGAN v2 \cite{choi2020stargan}&
0.329& 0.188 &
0.907 (0.005) &
0.552 (0.041) &
0.521 (0.036) &
0.589 (0.051) &
0.876 (0.009) &
0.502 (0.034) 
\\
SCCGAN \cite{guo2022alleviating}  &
0.321& 0.280&
0.842 (0.024) &
0.592 (0.043)&
0.392 (0.041)&
0.598 (0.043)&
0.892 (0.017)&
0.589 (0.035)
  \\
  DDIB \cite{su2022dual}&
0.339 & 0.205&		
 0.889 (0.006)&
 0.565 (0.027)&
 0.527 (0.024)&
 0.682 (0.026)&
 0.915 (0.009)&
 0.574 (0.021)\\

\cmidrule(lr){1-3} \cmidrule(lr){4-9}
\textbf{Ours}&
0.304& 0.296 &
\textbf{0.946} (0.012) &
\textbf{0.750} (0.042) &
\textbf{0.622} (0.048) &
\textbf{0.837} (0.038)&
\textbf{0.959} (0.008) &
\textbf{0.280} (0.039) \\
		\bottomrule[1pt]           
	\end{tabular}
	\end{center}
\label{tableMRI}
\end{table*}

\begin{figure*}[!h]
    \centering
\vspace{-10pt}
    \includegraphics[width =  0.99\textwidth]{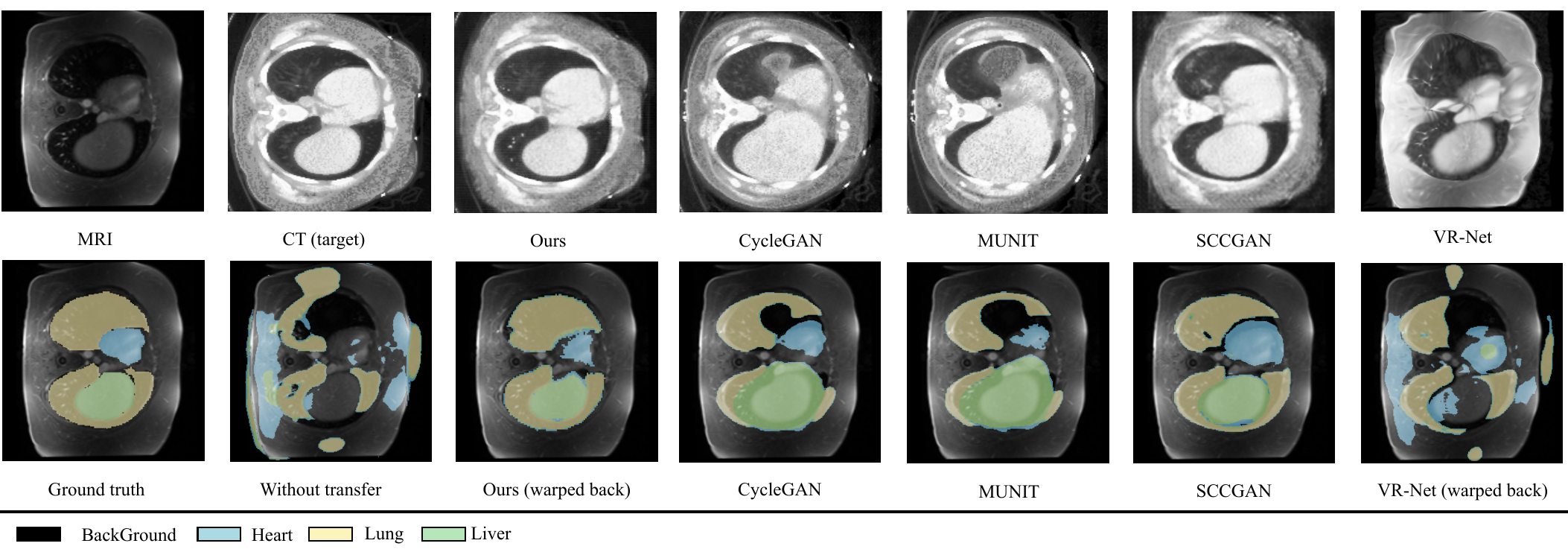}
    \caption{Results of translated images in the MRI to CT experiment.}
    \label{fig:MRI}
\end{figure*}

	As shown in Fig.~\ref{fig:OCT_trans} and Table \ref{tableOCT}, most GAN based methods could achieve good results in translating the intensity distributions. However,  MUNIT  cannot preserve the content information as it tends to change the structure to fit the target distribution, which seriously reduces the segmentation performance.   CycleGAN achieves balanced results in shifting the intensities and keeping structural information, however, some unnatural shapes and structures appear. Although spatial transformation methods reduce the shape gap between source and target data, the structure differences still persist to a certain extent and the intensity difference cannot be conquered.

\subsubsection{Ablation Study and Discussion}
\label{combination}

In order to justify each component of the proposed method, we conduct the following ablation studies.
The following methods are compared: (1) The baseline approach is the direct use of  the model trained on SINA   without any other processing, denoted as w/o Translation. (2) The proposed diffeomorphic spatial transformation  method to warp ATLANTIS images to SINA, denoted as DiffR. (3) The proposed DiffR method with clustering map  $u^*$, denoted as DiffR + $u^*$. (4) A combination of spatial transformation and structure-preserving GAN to the DiffR spatial transformation results, denoted as Transformation + GAN. (5) The combination in (4) with $u^*$, denoted as Transformation + GAN + $u^*$. (6) Our prpoposed method without the maximum SSIM mechanism. (7) The intensity translation module of the proposed method is replaced by the Fourier-transform-based method \cite{yang2020fda}. (8) Our proposed method. The results show that all the components are effective, as shown in
TABLE \ref{table:ablation}.   

Here we find out the reason that a straightforward combination of proposed spatial transformation method and the structure-preserving GAN cannot work well. It is because that the texture distortion caused by the spatial transformation is hard to be removed by the structure-preserving translation GAN. 	 
Fig. \ref{fig:ablation} shows an example for visual comparison. As shown, if we only employ a spatial transformation method, we are unable to obtain an accurate deformation. However, there are still many distortions in the translated results, such as the line-shape noise patterns. Such  distortions could lower the segmentation performance and can hardly be eased by intensity distribution translation methods. For example, source image warped by DiffR Attention and transferred intensity distributions still remains distorted. Because it is hard to separate these distortions from structure,  the network might consider them as structures rather than intensity distribution. Additionally, Fig. \ref{fig:Fourier} shows that the Fourier-transform-based method can not be effectively applied to replace the intensity translation module. Although Fourier-transform-based methods demonstrate notable proficiency in efficiently translating images, these methods are susceptible to introducing distortions in the translated results due to deformations. This susceptibility arises from the sensitivity of high-frequency components to deformations. Particularly, during the intensity translation phase, where inputs comprise deformed source images and basic structure contours, the high-frequency components of these deformed source images often contain distortions resulting from the deformations.

\subsection{Domain Translation from MRI to CT}

	\subsubsection{Datasets} To further justify the generalization of the proposed method, we also apply SUA on multi-modal data. In the second set of experiments, we apply it on domain transfer from MRI to CT. Two publicly available datasets\footnote{\url{https://learn2reg.grand-challenge.org}} \cite{leow2007statistical, kabus2009evaluation}   with manually labelled segmentation ground truth are used. The first dataset is a  MRI dataset that includes three regions in the chest, \textit{i.e.,} cardiac, lung, and liver. The second dataset is a CT dataset, taken from the same patients as the MRI dataset and containing the same organs. We use the MRI dataset as the  source, $X_S$, and the CT dataset as target, $X_T$. In this experiment, we only use the common areas of 2D images to  evaluate the effectiveness of the proposed method for its generalization.

	Cross modality translation often faces more comprehensive gaps in structure and intensity distribution. In some situations, the boundaries in the images can be fundamentally different because of the difference in imaging principles. For example, we observe  gradually changing brightness from the center to the top/bottom in MRI, whereas, a close to uniform distribution is observed in CT. In addition, shape is more flat in MRI than CT. 
 \begin{table*}[!t]
	\normalsize
	\setlength{\tabcolsep}{0.75 mm}
	\caption{Distribution and Segmentation Evaluation of Cardiac Images}
	\begin{center}
	\begin{tabular}{lp{1.2cm}p{1cm}p{1.9cm}p{1.9cm}p{1.9cm}p{1.9cm}p{1.9cm}p{1.9cm}}
	    \toprule[1pt]
		\multirow{2}{*}{Methods} & \multicolumn{2}{c}{Distribution} & \multicolumn{6}{c}{Segmentation} \\ 
        \cmidrule(lr){2-3} \cmidrule(lr){4-9}
		& $D_{Bhat}$ $\downarrow$ & $Corr$   $\uparrow$  &$Acc \uparrow$ & $Dice \uparrow$ & $mIoU \uparrow$ &  $Sen \uparrow$ & $Spe \uparrow$ & $FDR \downarrow$  \\
		\cmidrule(lr){1-3} \cmidrule(lr){4-9}
		Without Translation &
		
 0.204& 0.528&
0.956 (0.012) &
0.253 (0.036) &
0.237 (0.027) &
0.259 (0.027) &
0.755 (0.011) &
0.599 (0.251) 
\\

		VoxelMorph \cite{dalca2018unsupervised} &
0.153 &0.748 &
0.957 (0.012) &
0.298 (0.097) &
0.267 (0.067) &
0.290 (0.072) &
0.767 (0.029) &
0.511 (0.281) 
\\
		VoxelMorph+MI \cite{dalca2018unsupervised} &
 0.143& 0.755&
\textbf{0.959} (0.012) &
0.341 (0.142) &
0.299 (0.105) &
0.335 (0.128) &
0.786 (0.050) &
\textbf{0.504} (0.213) 
\\
		VR-Net \cite{jia2021learning} &
		0.144&0.776&
\textbf{0.959} (0.012) &
0.335 (0.118) &
0.292 (0.084) &
0.323 (0.098) &
0.781 (0.037) &
0.581 (0.185)    \\
		CycleGAN \cite{zhu2017unpaired}  &0.109& 0.881 &
0.958 (0.011) &
0.589 (0.127) &
0.399 (0.097) &
0.577 (0.221) &
\textbf{0.979} (0.010) &
0.599 (0.136) 
\\
		MUNIT \cite{huang2018multimodal} &
\textbf{0.059}&\textbf{0.960} &
0.946 (0.010) &
0.510 (0.067) &
0.333 (0.046) &
0.508 (0.193) &
0.969 (0.005) &
0.593 (0.049) 
\\
		DualGAN \cite{yi2017dualgan} &
		0.259& 0.825&
0.954 (0.012) &
0.275 (0.051) &
0.249 (0.034) &
0.272 (0.035) &
0.761 (0.016) &
0.561 (0.181)
\\
		NAGAN \cite{zhang2019noise}&
0.122& 0.902 &
0.956 (0.012) &
0.271 (0.057) &
0.248 (0.037) &
0.270 (0.038) &
0.759 (0.016) &
0.538 (0.228)
\\
		StarGAN v2 \cite{choi2020stargan} &
		0.185& 0.735&
0.944 (0.008) &
0.539 (0.066) &
0.353 (0.050) &
0.521 (0.150) &
0.966 (0.006) &
0.578 (0.053)  
\\

\textcolor{black}{ SCCGAN} \cite{guo2022alleviating} &
 0.117& 0.940&
0.950 (0.010)&
0.553 (0.096)&
0.367 (0.074)&
0.605 (0.137)&
0.939 (0.045)&
0.553 (0.083)
\\
		DDIB \cite{su2022dual}&
0.114 & 0.940 &
 0.946 (0.012)&
 0.236 (0.003)&
 0.223 (0.006)&
 0.244 (0.004)&
 0.745 (0.003)&
 0.739 (0.113)\\
\cmidrule(lr){1-3} \cmidrule(lr){4-9}
\textbf{Ours} & 
0.116&  0.948&
0.953 (0.025) &
\textbf{0.541} (0.221) &
\textbf{0.564} (0.240) &
\textbf{0.633} (0.209) &
0.903 (0.073) &
0.556 (0.215) 
 \\
		\bottomrule[1pt]           
	\end{tabular}
	\end{center}
	\label{tablecardiac}
\end{table*}

\begin{figure*}[!t]
    \centering
\vspace{-10pt}
    \includegraphics[width = \textwidth]{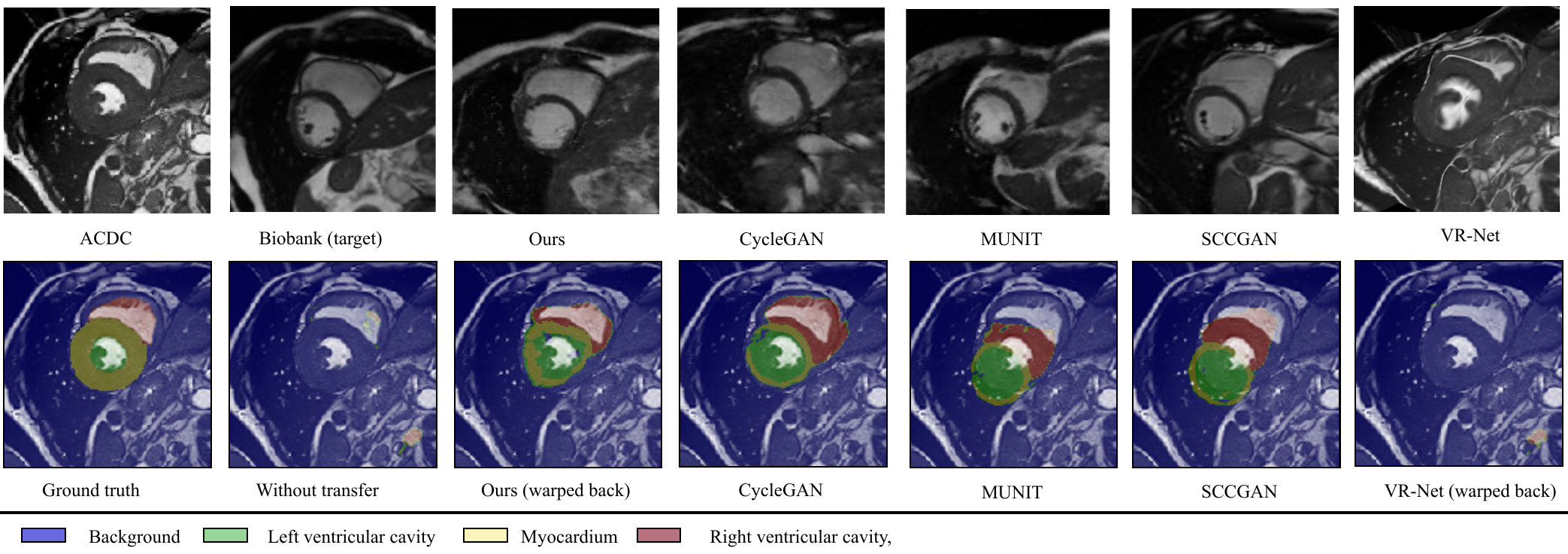}
    \vspace{-15pt}
    \caption{Results of translated images in the ACDC to UKBB experiment.}
    \label{fig:cardiac_trans}
\end{figure*}
 
	\subsubsection{Comparison with prior arts} 
	Similar to retinal OCT experiments, we train the segmentation network on CT and use it to infer transferred MRI images. Similarly, the segmentation outputs are warped back and evaluated against the original MRI ground truth. 
 We   compare the proposed method with the same set of methods as in Section \ref{octexp}.
 
	\textbf{Evaluation of  Structure Transfer}: We first evaluate the structure transfer from MRI to CT.
	 As shown in the TABLE \ref{tableMRI} and Fig. \ref{fig:MRI}, our proposed SUA method outperforms the  state-of-the-art methods on $mIoU$, $Dice$, $Acc$, $Spe$ and $FDR$. Structures   of cardiac, lung and liver are transferred successfully compared with the VR-Net, Voxelmorph and DualGAN methods. While our inverse deformation can maintain the original shapes compare with diffusion based translation model DDIB and translation GANs such StarGAN v2, MUNIT, NAGAN, \textcolor{black}{SCCGAN} and CycleGAN.

\textbf{Evaluation of  Intensity Distribution Transfer}: TABLE \ref{tableMRI} compares the proposed method with other methods in transferring intensity distribution. Fig.~\ref{fig:MRI} gives some examples for visual comparison.  As shown in the results, translation GANs could also obtain good results on translating the intensity distributions similar to the OCT experiment. Unsurprisingly, MUNIT achieves the best distribution results but obtains bad results in segmentation. This is reasonable since MUNIT changes the semantic information and leads to misaligned masks. Furthermore, we can see that the translated images by our method not only reach good results in distribution alignment but also capture the structure information by obtaining the inverse deformations.
	Although spatial transformation methods reduce the shape gap between source and target datasets, the structure difference still remains to a certain extent and the intensity difference cannot be conquered.

\subsection{Translation between Data from Healthy and Unhealthy Subjects}

   \subsubsection{{Datasets}}
   Next, we examine how  our method performs when translating between data from healthy subjects and that from unhealthy subjects. For this we use MRI images from both the ACDC \cite{bernard2018deep} and the Biobank (UKBB) \cite{petersen2013imaging} datasets.  For ACDC, it is composed of 100 patients with three types of pathologies: infarction, dilated cardiomyopathy, and hypertrophic cardiomyopathy. For UKBB, we randomly select 100 healthy subjects. The aim is to transfer the style of images in ACDC to that in UKBB, so that the segmentation model trained from healthy patients also works for pathological cases. For each subject in both datasets, we select images at the end-diastolic (ED) frames for experiments.


	
	\subsubsection{Translation from Unhealthy Subjects Data to Healthy Subjects Data}
    Similar to the retinal OCT experiments and the MRI-to-CT experiments, we train a segmentation model on normal images in UKBB which is then used to segment the diseased images in ACDC dataset after image translation. We then evaluate the performance by comparing warped segmentation outputs and original ACDC ground truth. Similarly, the proposed method is compared with the same set of methods as in Section \ref{octexp}.

    \textbf{Evaluation of  Structure Translation}:
    TABLE \ref{tablecardiac} shows the performance of different models measured by $mIoU$, $Dice$, $Acc$, $Spe$ and $FDR$. Compared with the VR-Net, Voxelmorph, and DualGAN methods, our SUA model obtains better results. Additionally, Figure. \ref{fig:cardiac_trans} shows that the transferred structures are on par with that by VoxelMorph, VoxelMorph(MI), and VR-net. Furthermore, the inverse deformation obtained by our method can maintain the shape information of the source dataset, thus outperforming diffusion based translation model DDIB and translation GANs, such as CycleGAN, NAGAN, \textcolor{black}{SCCGAN} and StarGAN v2.

		\textbf{Evaluation of  Intensity Distribution Transfer}: 
	Similarly, we evaluate translation on intensity distributions. TABLE \ref{tablecardiac} shows the comparison between the proposed method and other methods. Specifically, translation GANs have advantages in the translation of intensity distribution, which has been observed similarly in previous experiments. Not unexpectedly, the same issues occur in MUNIT: though it gives the best distribution results visually, it loses the semantic information of the input source samples. In contrast, the translated images by our method not only produce good results in aligning distributions, but also capture the structure information by using the inverse deformations. Spatial transformation methods, such as VoxelMorph with mutual information loss, reduce the structure gaps to a certain extent, but the intensity difference still could not be decreased.

 \subsubsection{Translation from Healthy Subjects Data to Unhealthy Subjects Data}
\begin{table}[h!]
    \centering
      \caption{Translation measurement on reverse direction.}%
    \resizebox{\linewidth}{!} { 
    \begin{tabular}{l||c|c}
			\toprule[1.5pt]
		Method  
		& $D_{Bhat}$ $\downarrow$ & $Corr$   $\uparrow$   \\\hline
		without Translation &0.204 & 0.528\\
		Reverse direction (health to disease) & \textbf{0.107}& \textbf{0.901} \\
		\bottomrule[1pt]         
	\end{tabular} \label{table:reverse}
 }

\end{table}
\begin{figure}[h!]
    \centering
 \includegraphics[width=\linewidth]{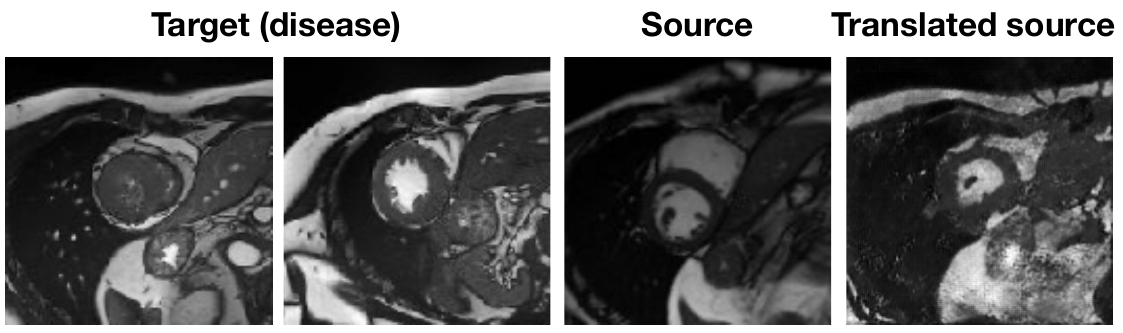}
}{%
  \caption{Health to disease translation.}%
    \label{fig:reverse}
\end{figure}
 
To further understand the relationship between normal and abnormal image characteristics, we conducted an additional experiment in the reverse direction.  This involved translating images representative of a healthy state into the distribution typically associated with disease images. The results, as depicted in the subsequent TABLE  \ref{table:reverse} and Fig. \ref{fig:reverse}, the gaps between normal cases and abnormal cases are reduced by translation. Specifically, the diseases in these cardiac samples mainly manifest as morphological and structural abnormalities in images, rather than the presence of additional lesions, which makes generating images from healthy to disease not out of reach.



\section{Discussion}

\begin{figure*}[t!]
    \centering
    \includegraphics[width=\textwidth]{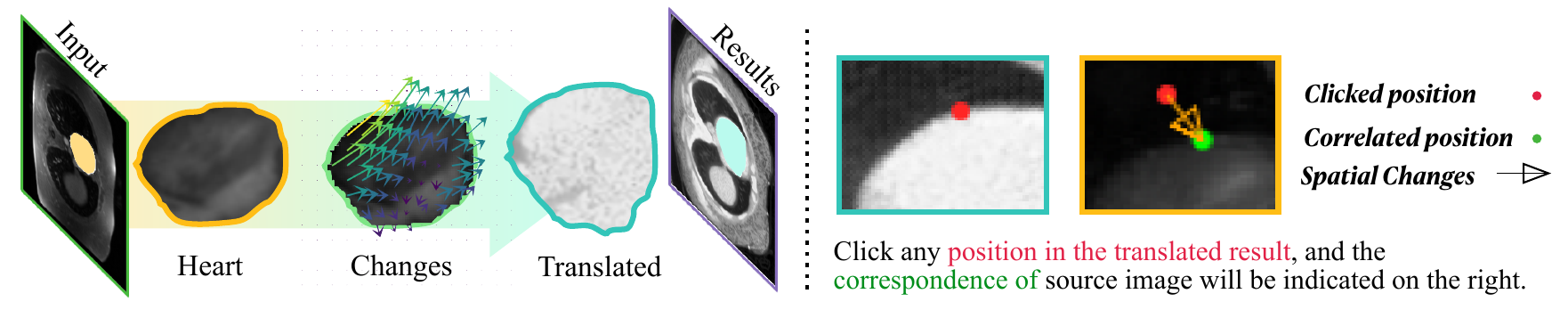}
    \caption{\textcolor{black}{Illustrating the translation traceability. Left: a semantic figure showing how the heart is deformed and translated from MRI to CT. Right: a screenshot of the online demo given at \url{https://traceable-translation.github.io}) showing the spatial connection before and after translation.}} 
    \label{fig:demo_app}
\end{figure*}

 \begin{figure}[h!]
    \centering
    \includegraphics[width = 0.8\linewidth]{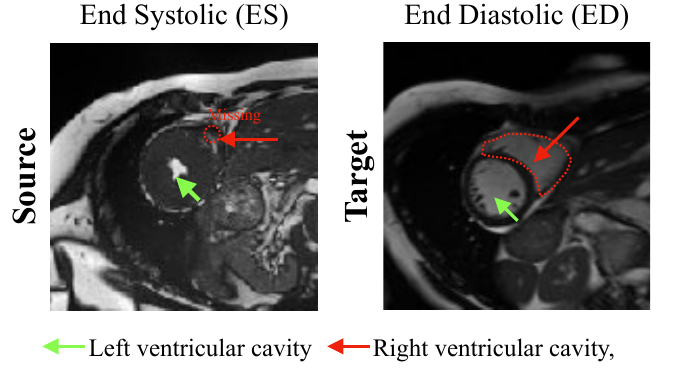}
    \caption{\textcolor{black}{The first image is one of source images, which is an abnormal image  on ES state; the second image is a target image which is normal and is on ED state.}}
    \label{fig:demo_f}
    \vspace{-10pt}
\end{figure}   

\begin{table*}[t!]
\centering
    \caption{The Mutual Information (Median) between source and target datasets (source \& target) and between source and translated source datasets (source \& reuslts).}
\resizebox{\textwidth}{!}{%
        \begin{tabular}{lcc cc cc}
        \toprule[1pt]
        & \multicolumn{2}{c}{\textbf{ Retinal OCT}} & \multicolumn{2}{c}{\textbf{Chest MRI \& CT}} & \multicolumn{2}{c}{\textbf{Cardiac MRI}} \\
        \cmidrule(lr){1-3} \cmidrule(lr){4-5} \cmidrule(lr){6-7}
       
    \textbf{Structual Gap $\downarrow$}	&source \& target & source \& results &source \& target & source \& results &source \& target & source \& results \\
    \midrule
        		Mutual Information $\uparrow$&
 0.0317&
0.1247 &
 0.3072 &
 0.9956 &
0.1167&
0.2186 \\

\bottomrule
\label{Mutual_inf}
 \end{tabular}   
    }
    \label{tab:seg_res_table}
    \vspace{-10pt}
\end{table*}

\subsection{Quantitative Investigation}
We have rigorously evaluated the proposed method across six different datasets of varied modalities, extensive experimental results have also proved the superiority of the proposed method in MRI-CT translation and pathology-normal cardiac MRI translation. In these two tasks, large structural gaps can be clearly observed. These results from the six datasets clearly prove that our method is a general approach and can effectively translate images between various domains and reduce the shifts to improve segmentation with large and small structural gaps. As quantification of the extents of structure gap could be important, we employ Mutual Information loss. This metric is adept at assessing structural differences and has been effectively used to measure structural consistency post-translation in SCCGAN \cite{guo2022alleviating}. A lower Mutual Information loss between two datasets indicates a greater structural divergence. Our computations reveal that the initial Mutual Information loss between OCT datasets is substantial but is significantly mitigated by our translation method, as evidenced in the translated results shown in TABLE \ref{Mutual_inf}.

We noticed that large variances appear in the results of the third dataset, \emph{e.g}, Ours and CycleGAN's. 
This is because this cardiac dataset is a very difficult one and these challenges come from data collected at different phases and translation from pathological to healthy subjects. Specifically, the source images in this dataset are from abnormal subjects at the end systolic (ES) phase, whilst the target images are from normal subjects at the end diastolic (ED) phase. As shown in Fig. \ref{fig:demo_f} left, the left ventricular cavity in the source image is very small, and the right ventricular cavity even disappears. In contrast, they are completely normal in the target image. These issues make it challenging to estimate the deformation between them, which led to large variances in our final results. This high variance problem can be migrated by either using large-sized datasets (like our first OCT dataset) or datasets that have smaller structure gaps (like our second MR/CT dataset). We highlight that although our method produced higher variances in this cardiac dataset, our method still achieved the best overall results compared to other baseline methods. The superiority of our method can be confirmed by its higher means in TABLE \ref{tablecardiac}.
\color{black}

\subsection{\textcolor{black}{Structure Correspondence}}
\label{structure}
 The significance of structure correspondence is often overlooked in existing image-to-image methods. They usually focus on whether the generated image are similar to the target distribution rather than whether the structure of the generated image is related to that of the source image \cite{guo2022alleviating}. Furthermore, preserving the structure without addressing the geometry gaps between source and target images is inadequate for fulfilling the translation requirements in medical image tasks. For example, the geometry difference between the imaging  in MRI to CT translation  would affect the treatment course \cite{Emilia2021}. It is also mentioned in \cite{hering2022learn2reg} that establishing correspondences can help to monitor disease progression, estimate motion in radiotherapy planning. These imply that structure correspondence is essential for treatment courses. 

Our proposed method possesses an ability to synthesize large quantities of paired data by employing structure unbiased image-to-image translation, while achieving pixel-level structure correspondence. This breakthrough paves a way for the development of an interactive translation method, enabling accurate translation of images with point-to-point correspondences to effectively capture structural changes. To exemplify this advancement, we presented Fig. \ref{fig:demo_app} to show the spatial correspondences between source and translated images. An \textbf{\textit{online demo}} was also developed and given at \url{https://traceable-translation.github.io}.

\section{Conclusion}

Image to image translation is an essential task in machine learning, especially for tasks across different modalities.  In this paper, we propose to reduce the domain shifts in both  structure and intensity distributions. A novel  SUA network which contains a structure extractor, a spatial transformation module, and an intensity-rendering module, is proposed in this paper. Our experimental results have shown that SUA is able to transfer both the structure and intensity distributions and improve segmentation results.
Although our method achieves state-of-the-art performance, it has a limitation over the domain gaps introduced by  very small objects or lesions as they are often too small to be detected by the dominant structure extractor, which would be  future work.


%




\section{Acknowledgments}
We highly acknowledge the anonymous reviewer for his/her contribution to improving the quality of this manuscript.

The computations described in this research were performed using the Baskerville Tier 2 HPC service. Baskerville was funded by the EPSRC and UKRI (EP/T022221/1 and EP/W032244/1) and is operated by Advanced Research Computing at the University of Birmingham. This research is partially funded by the BHF Accelerator Award (AA/18/2/34218) and by the Korea cardiovascular Bioresearch Foundation (CHORUS Seoul 2022). This research used the UK Biobank Resource under the application number 40119. This work is also partially supported by Huazhu Fu’s A*STAR Central Research Fund, and AI Singapore Programme (No: AISG2-TC-2021-003).

\bibliographystyle{IEEEtran}
\bibliography{egbib.bib}
\begin{IEEEbiography}[{\includegraphics[width=1in,height=10.25in,clip,keepaspectratio]{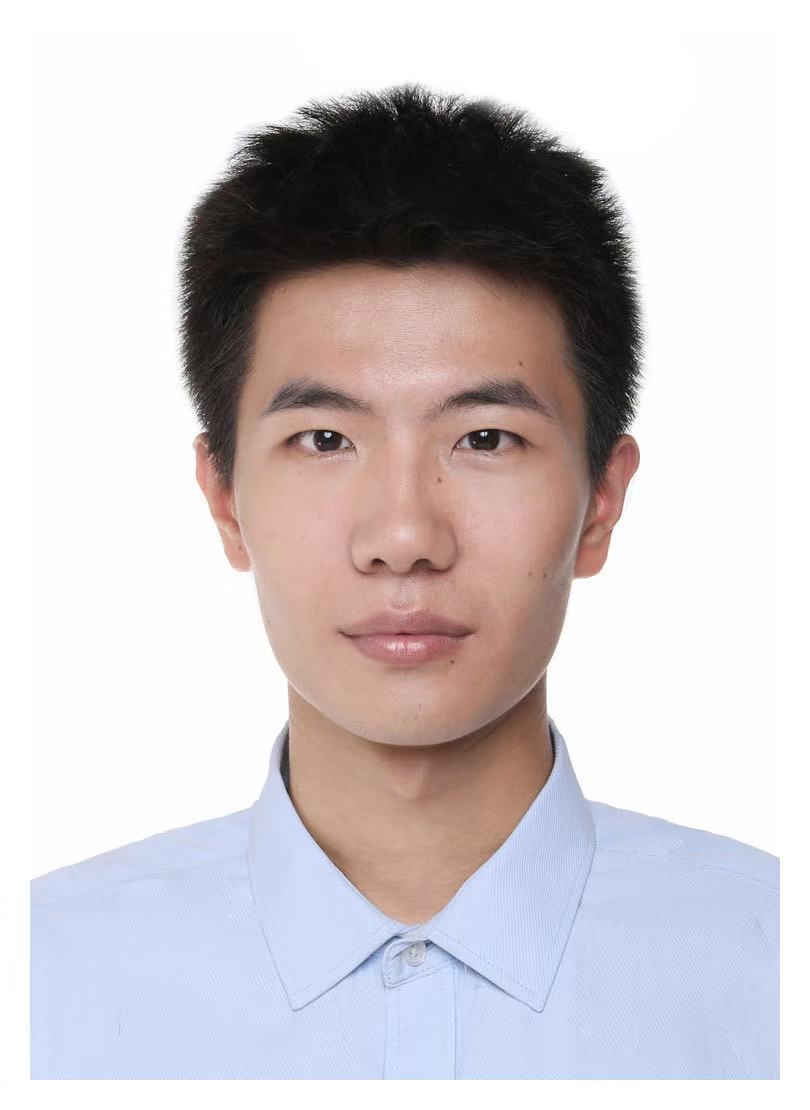}}]{Tianyang Zhang} is now a PhD. candidate of the school of computer science at the university of Birmingham, UK. Advised by Jinming Duan and Aleš Leonardis. His research interests include, but not limited to, medical image translation, and medical image registration.
\end{IEEEbiography}
\vspace{-30pt}
\begin{IEEEbiography}[{\includegraphics[width=1in,height=10.25in,clip,keepaspectratio]{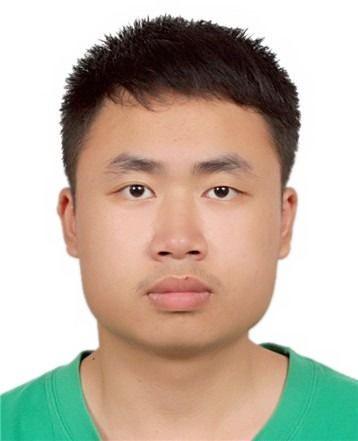}}]{Shaoming Zheng} is now a Ph.D. student of the Department of Electrical and Electronic Engineering at Imperial College London, London, UK. His research interests include, but not limited to, medical image analysis and synthesis.
\end{IEEEbiography}
\vspace{-30pt}
\begin{IEEEbiography}[{\includegraphics[width=1in,height=10.25in,clip,keepaspectratio]{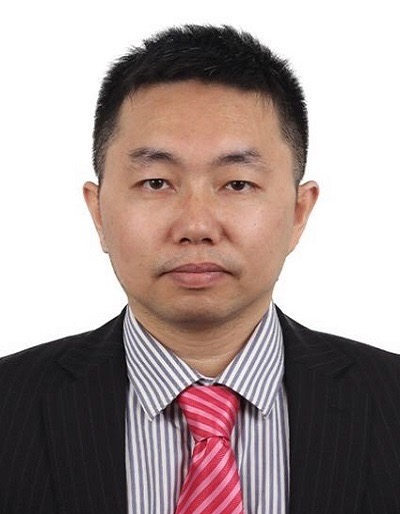}}]{Jun Cheng}(SM’20) received the B. E. degree in electronic engineering and information science from the University of Science and Technology of China, and the Ph. D. degree from Nanyang Technological University, Singapore. He is now a senior research scientist in the Institute for Infocomm Research, Agency for Science, technology and Research,  working on AI for medical imaging, robust machine vision and perception. He is an Associate Editor for IEEE TIP and IEEE TMI. 
\end{IEEEbiography}
\vspace{-30pt}
\begin{IEEEbiography}[{\includegraphics[width=1in,height=10.25in,clip,keepaspectratio]{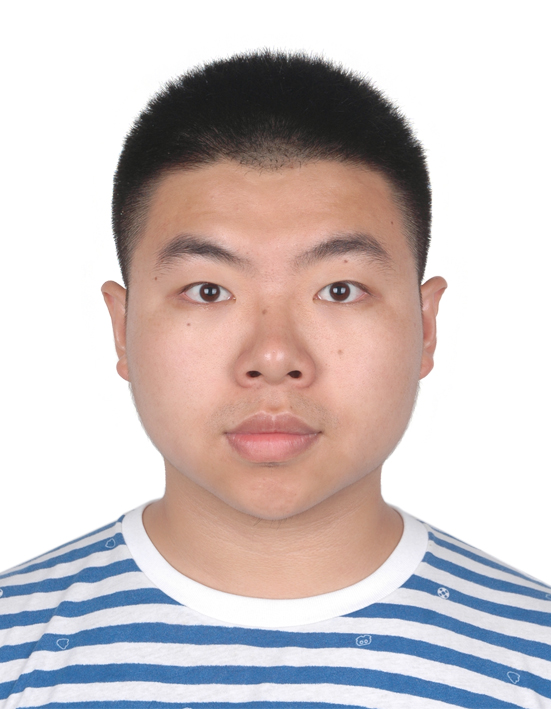}}]{Xi Jia}
is a Ph.D. candidate at the University of Birmingham. His research interests include pattern recognition and medical image processing.
\end{IEEEbiography}
\vspace{-30pt}
\begin{IEEEbiography}[{\includegraphics[width=1in,height=10.25in,clip,keepaspectratio]{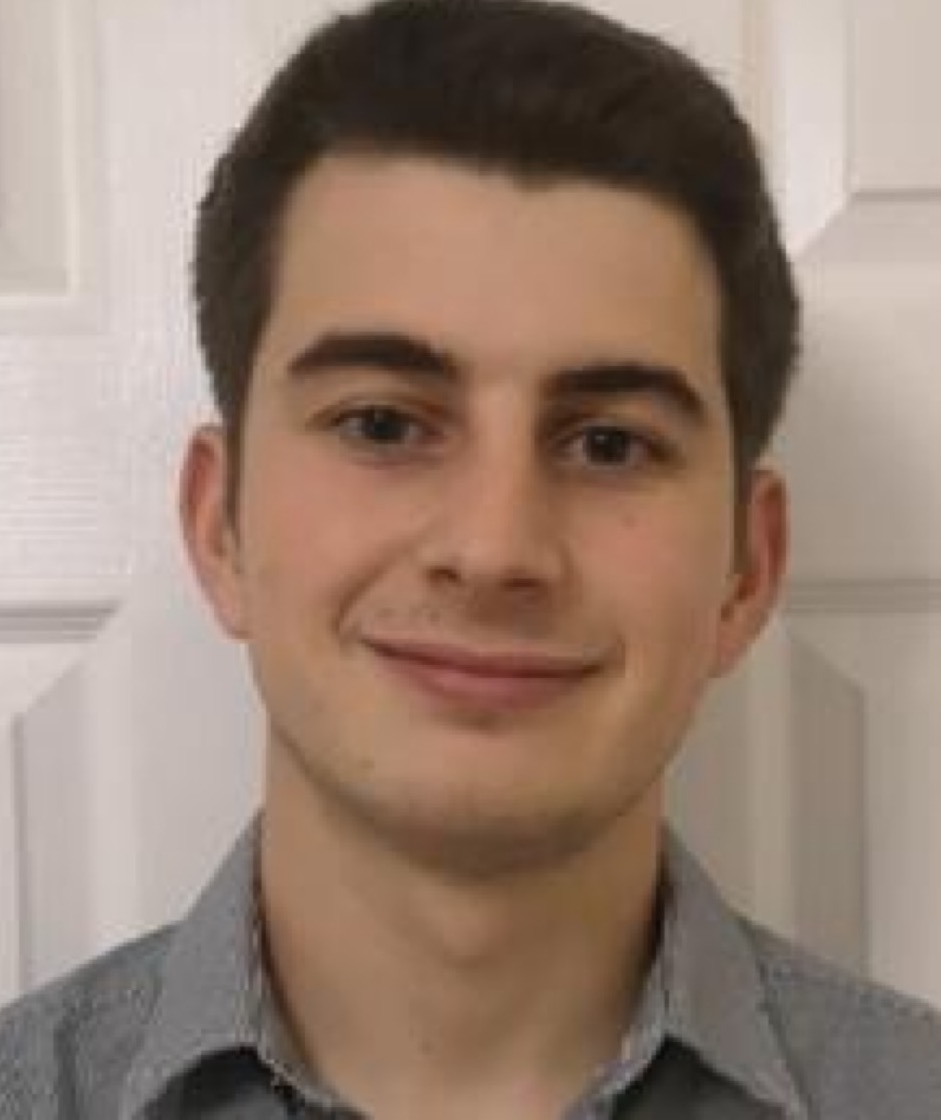}}]{Joseph  Bartlett}
 is a PhD candidate studying jointly at the University of Birmingham and the University of Melbourne, his research focuses on improving diffusion magnetic resonance imaging using machine learning techniques.
\end{IEEEbiography}
\vspace{-30pt}
\begin{IEEEbiography}[{\includegraphics[width=1in,height=10.25in,clip,keepaspectratio]{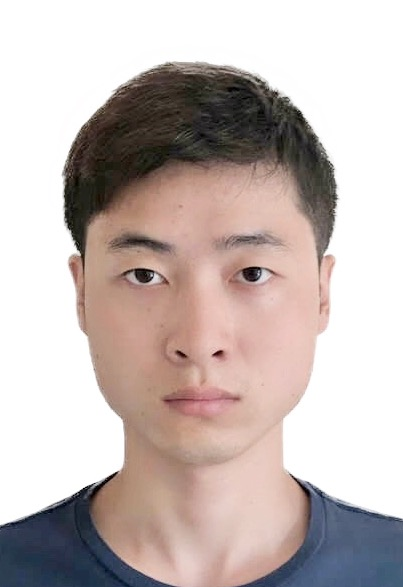}}]{Xinxing Cheng}
 is a PhD student of the school of computer science at the university of
Birmingham, UK. His research interests include, but not limited to, medical image analysis and synthesis.
\end{IEEEbiography}
\vspace{-30pt}
\begin{IEEEbiography}[{\includegraphics[width=1in,height=10.25in,clip,keepaspectratio]{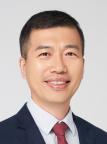}}]{Zhaowen Qiu} is the Director of China Computer Society (CCF), Outstanding member of CCF, Standing member of CCF computer application Committee, Member of Computing Machinery (ACM), Institute of Electrical and Electronics Engineers (IEEE), Artificial Intelligence Committee of China Anti Cancer Association (CACA), Digital diagnosis and treatment Committee of Geriatric Society.Director of Institute of 3D digital technology, Northeast Forestry University.Director of Heilongjiang medical image 3D visualization and 3D printing engineering technology center.
\end{IEEEbiography}
\vspace{-30pt}

\begin{IEEEbiography}[{\includegraphics[width=1in,height=10.25in,clip,keepaspectratio]{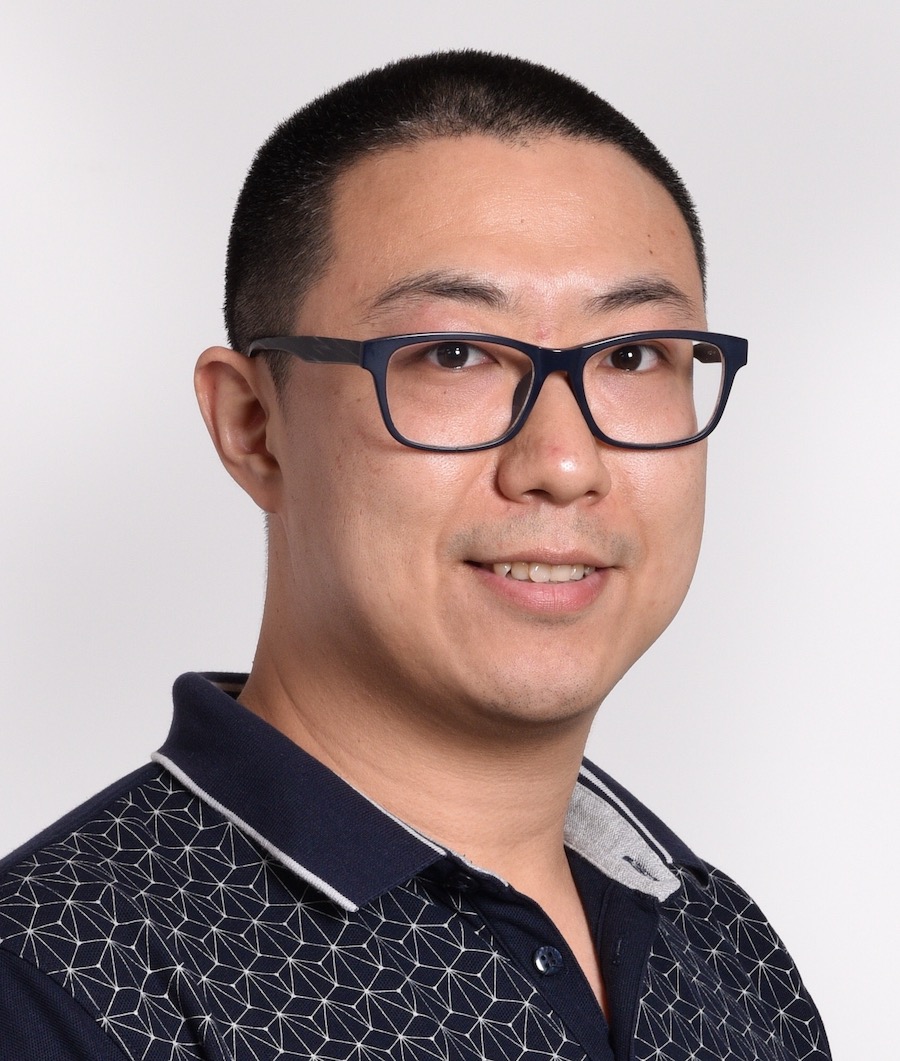}}]{Huazhu Fu} (SM'18) is a senior scientist at Institute of High Performance Computing (IHPC), A*STAR, Singapore. He received his Ph.D. from Tianjin University in 2013. Previously, he was a Research Fellow (2013-2015) at NTU, Singapore, a Research Scientist (2015-2018) at I2R, A*STAR, Singapore, and a Senior Scientist (2018-2021) at Inception Institute of Artificial Intelligence, UAE. His research encompasses computer vision, AI in healthcare, and trustworthy AI. He has received a number of awards including Best Paper Award at ICME 2021, and Best Paper Award at MICCAI-OMIA 2022. He is an Associate Editor of IEEE TMI, IEEE TNNLS, IEEE TAI, and IEEE JBHI. He also serves on the Bio Imaging and Signal Processing Technical Committee (BISP TC) of the IEEE Signal Processing Society
\end{IEEEbiography}
\vspace{-30pt}
\begin{IEEEbiography}[{\includegraphics[width=1in,height=10.25in,clip,keepaspectratio]{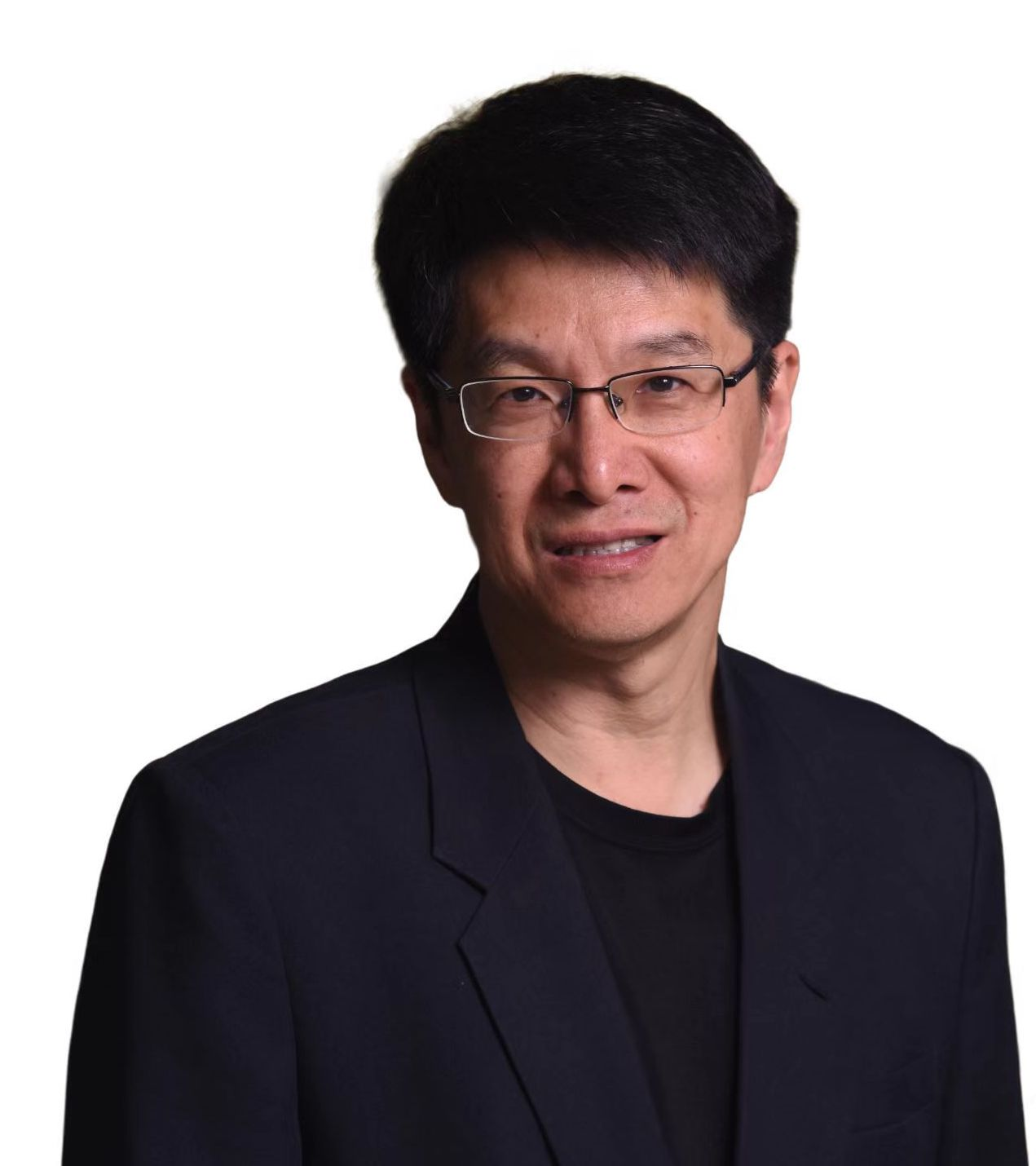}}]{Jiang Liu} is a full professor in the Department of Computer Science and Engineering at the Southern University of Science and Technology. He received an M.S. (1992) and a Ph.D. (2004) in Computer Science from the National University of Singapore. His main research interests include medical image processing and artificial intelligence.

\end{IEEEbiography}
\vspace{-30pt}
\begin{IEEEbiography}[{\includegraphics[width=1in,height=10.25in,clip,keepaspectratio]{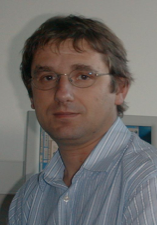}}]{Ale\v{s} Leonardis}
is a professor at the School of
Computer Science, University of Birmingham, the Chair
of robotics with the University of Birmingham, and the co-director of the Centre for Computational Neuroscience and Cognitive Robotics. He is also a professor at the Faculty of Computer
and Information Science, University of Ljubljana
and adjunct professor at the Faculty of Computer
Science, TU-Graz. His research interests include robust and adaptive
methods for computer vision, object and scene
recognition and categorization, and biologically
motivated vision.
\end{IEEEbiography}
\vspace{-30pt}
\begin{IEEEbiography}[{\includegraphics[width=1in,height=10.25in,clip,keepaspectratio]{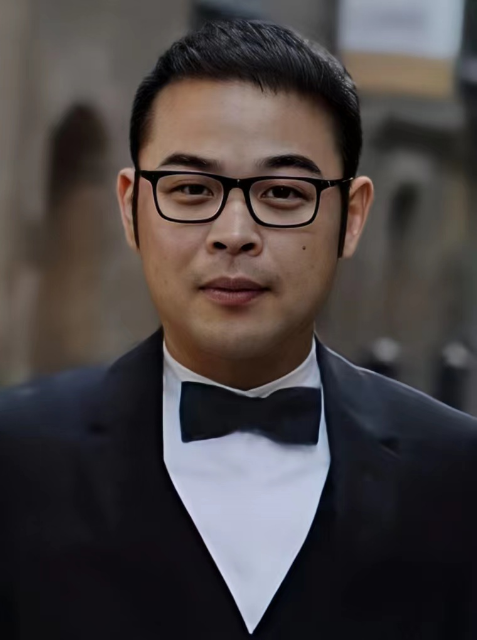}}]{Jinming Duan} is now a Turing Fellow at Alan Turing Institute and an Assistant Professor within Computer Science at University of Birmingham. He is also a Fellow of the Higher Education Academy (FHEA) under the UK Professional Standards Framework for teaching and learning support in higher education. He was a Research Associate jointly within Department of Computing \& Institute of Clinical Sciences at Imperial College London. There, he has been developed cutting-edge machine learning methods for cardiovascular imaging problems. Prior to that, he acquired his PhD degree in Computer Science at University of Nottingham. His PhD was funded by Engineering and Physical Sciences Research Council. Jinming’s research includes deep neural nets, variational methods, partial/ordinary differential equations, numerical optimisation, and finite difference/element methods, with applications to image processing, computer vision and medical imaging analysis. 
\end{IEEEbiography}

\end{document}